\documentclass{aa}  

\usepackage{graphicx}
\usepackage{txfonts}
\usepackage{hyperref}
\begin{document}

   \title{Clustered star formation towards Berkeley 87 / ON2}

   \subtitle{I. Multi-wavelength census and the population overlap problem}

   \author{Diego de la Fuente\inst{1,2,3}
          \and
          Carlos G. Rom\'an-Z\'u\~niga\inst{3}
          \and
          Elena Jim\'enez-Bail\'on\inst{3}
          \and
          Jo\~ao Alves\inst{4}
          \and
          Miriam Garcia\inst{2}
          \and
          Sean Venus\inst{3}
          }

   \institute{Departamento de F\'isica, Ingenier\'ia de Sistemas y Teor\'ia de la Se\~nal, Universidad de Alicante,
Carretera de San Vicente s/n, 03690, San Vicente del Raspeig, Alicante, Spain\\
              \email{diego.delafuente@ua.es}
         \and
             Departamento de Astrof\'isica, Centro de Astrobiolog\'ia (CSIC-INTA), Ctra. Torrej\'on a Ajalvir km 4, 28850 Torrej\'on de Ardoz, Spain
         \and    
             Instituto de Astronom\'ia, Unidad Acad\'emica de Ensenada, Universidad Nacional Aut\'onoma de M\'exico, Ensenada 22860, Mexico
         \and
             Institute for Astrophysics, University of Vienna, Türkenschanzstrasse 17, 1180 Vienna, Austria
             }

   \date{Received December 4, 2020; accepted \today}

 
  \abstract
   {Disentangling line-of-sight alignments of young stellar populations is crucial for observational studies of star-forming complexes. This task is particularly problematic in a Cygnus-X subregion where several components, located at different distances, are overlapped: the Berkeley 87 young massive cluster, the poorly-known [DB2001] Cl05 embedded cluster, and the ON2 star-forming complex, in turn composed of several H\textsc{ii} regions.}
   {To provide a methodology for building an exhaustive census of young objects that can consistently deal with large differences in both extinction and distance.}
   {OMEGA2000 near-infrared observations of the Berkeley 87 / ON2 field are merged with archival data from \textit{Gaia}, \textit{Chandra}, \textit{Spitzer}, and \textit{Herschel}, as well as cross-identifications from the literature. To address the incompleteness effects and selection biases that arise from the line-of-sight overlap, we adapt existing methods for extinction estimation and young object classification, and we define the intrinsic reddening index, $R_\mathrm{int}$, a new tool to separate intrinsically red sources from those whose infrared color excess is caused by extinction. We also introduce a new method to find young stellar objects based on $R_\mathrm{int}$.}
   {We find 571 objects whose classification is related to recent or ongoing star formation. Together with other point sources with individual estimates of distance or extinction, we compile a catalog of 3005 objects to be used for further membership work. A new distance for Berkeley 87, $(1673 \pm 17)~\mathrm{pc}$, is estimated as a median of 13 spectroscopic members with accurate \textit{Gaia} EDR3 parallaxes.}
   {The flexibility of our approach, especially regarding the $R_\mathrm{int}$ definition, allows to overcome photometric biases caused by large extinction and distance variations, in order to obtain homogeneous catalogs of young sources. The multi-wavelength census that results from applying our methods to the Berkeley 87 / ON2 field will serve as a basis for disentangling the overlapped populations.}


   \maketitle
%

\section{Introduction} \label{sec:intro}

The closest giant star-forming complex, Cygnus-X, is a privileged site to observe the conditions under which massive star clusters are born and grown up. The rich variety of star-forming clouds, OB clusters and associations that are part of Cygnus-X \citep{leduigou-knodlseder02, reipurth-schneider08} is ideal for understanding how feedback from massive stars influence the formation of new clusters in the surroundings. To assess this interplay between components, it is crucial to know precisely the 3-dimensional structure of the complex. In other words, the young populations that are overlapped along the line of sight must be disentangled.

Measuring distances towards Cygnus-X is, however, a challenging task. The complex is located in a kinematic avoidance zone \citep{ellsworthbowers+15}, where the line of sight is nearly tangential to the Galactic rotation. This implies that kinematic distances are highly uncertain up to several kiloparsecs, to the point that deciding whether Cygnus-X components are physically connected or their distances are not even comparable has been a long-standing controversy \citep{pipenbrink-wendker88, uyaniker+01, schneider+06, rygl+12}.

With the advent of the \textit{Gaia} mission \citep{gaia+16}, the line-of-sight overlap problem can be partially solved. \textit{Gaia} parallaxes yield accurate distance results for optically visible stellar populations in Cygnus-X (\citealt{berlanas+19}; \citealt{lim+19}). Distances to dust clouds can be also derived from \textit{Gaia} parallaxes if reddened stars located behind them are still detectable in optical wavelengths \citep{zucker+19,zucker+20, alves+20}. These techniques, however, are no longer valid under heavier extinction conditions, and not applicable to individual point sources that are only detectable in longer wavelengths (e.g. masers, protostars).

\begin{figure*}
	\centering
	\includegraphics[height=8cm]{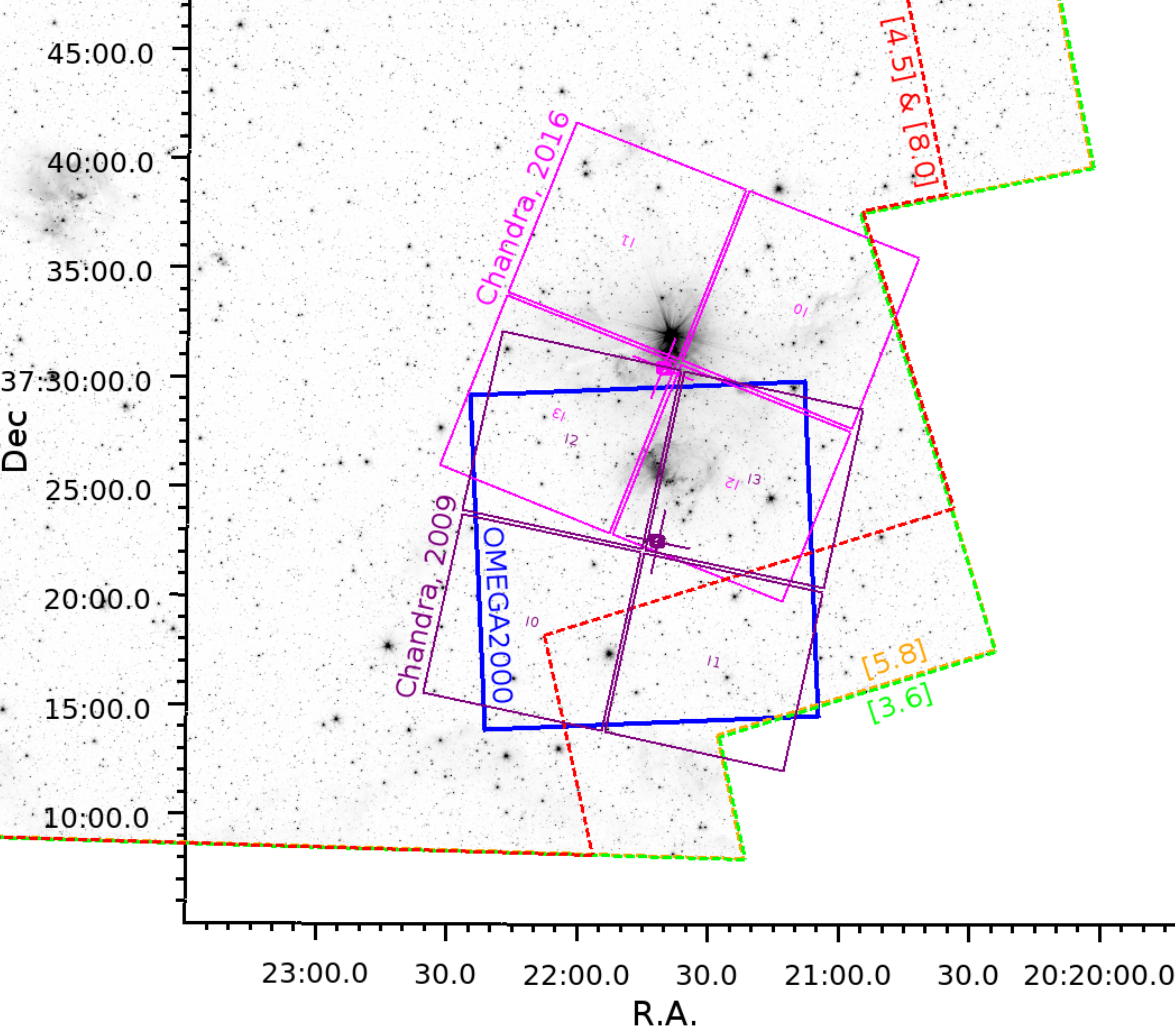}
	~
	\includegraphics[height=8cm]{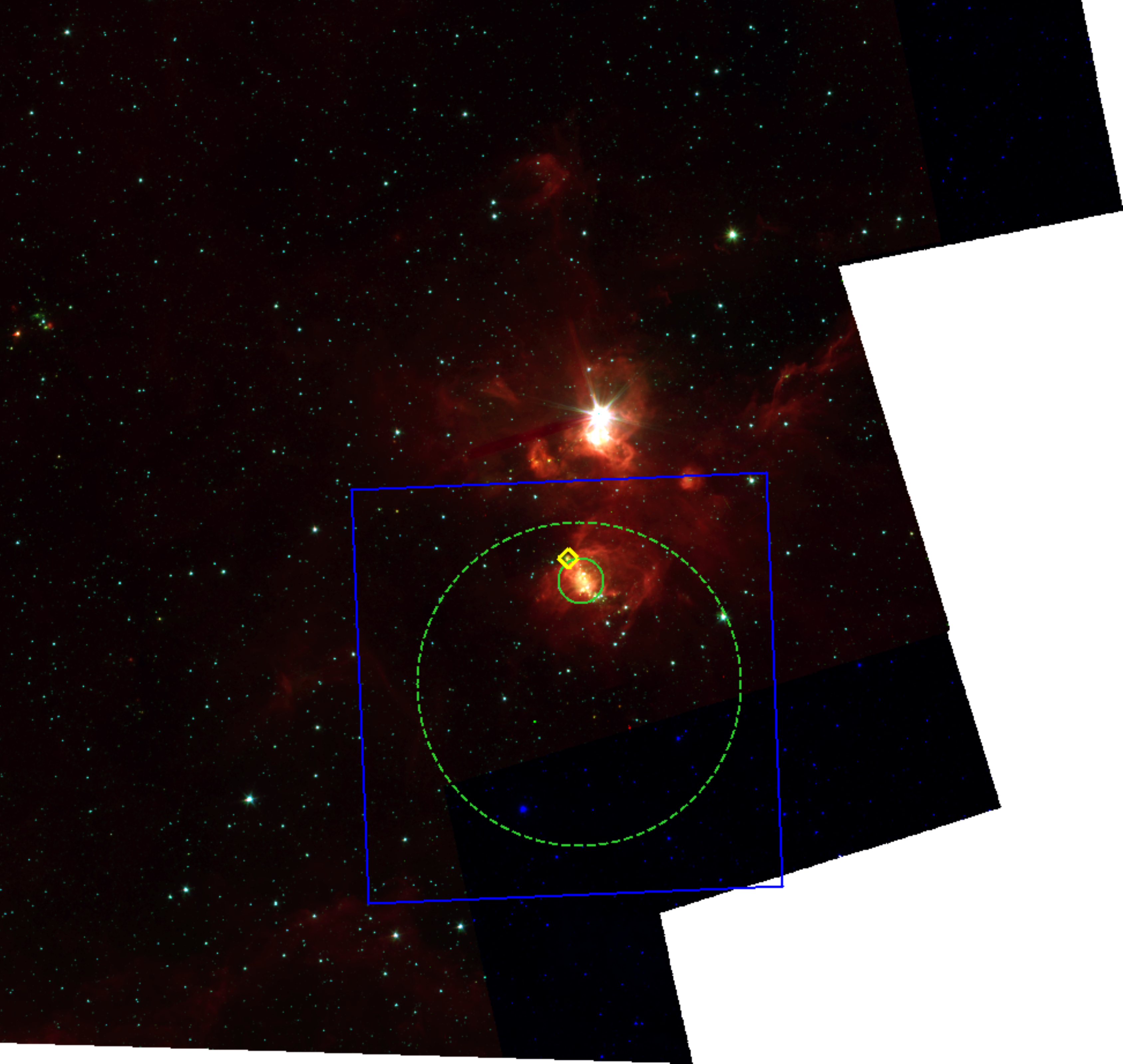}
	\caption{\textit{(Left)} Spatial coverage of the OMEGA2000, Chandra, and IRAC observations on the $3.6~\mu \mathrm{m}$ image of the ON2 region. \textit{(Right)} RGB image ($R=[8.0], G=[4.5], B=[3.6]$) of the region that is covered by coordinate axes in the left pannel, and showing the OMEGA2000 field of view in the same way. Green circles show the position and size of Berkeley 87 (dashed) and [DB2001] Cl05 according to \citet{dias+14} and \citet{dutra-bica01}, respectively. The Cygnus 2N region is marked as a yellow diamond. \label{fig:coverage}}
\end{figure*}

In this series of papers, we deal with a particularly troublesome subregion in southwestern Cygnus-X, where several components are projected onto the same field. The most relevant objects related with clustered star formation are introduced below (and shown in Fig. \ref{fig:coverage}).

Berkeley 87 is a young massive cluster, located $\sim 1.7~\mathrm{kpc}$ away, that is famous for hosting massive stars in rare phases of their late evolution, like the peculiar emission-line variable V439 Cyg \citep{polcaro+89,polcaro-norci98}. It is also the only cluster in the Milky Way that is currently known to host a WO (oxigen-rich Wolf-Rayet) star, WR 142 \citep{barlow-hummer82,drew+04,oskinova+09,rate+20}. The first complete characterization of Berkeley 87 was carried out by \citet{turner-forbes82}, and further studies were presented by \citet{polcaro+91,massey+01,turner+06,bhavya+07,majaess+08,sokal+10,oskinova+10}.

ON2\footnote{Historically, the ``ON2'' and ``Onsala 2'' denominations have been used interchangeably, either for the whole star-forming region, or its smaller constituents (H\textsc{ii} regions, masers). This is a source of confusion for the related nomenclature, e.g. ``Onsala 2N'' is \textit{not} related to ``ON 2N''. To make things clear, in this series of papers we will follow the \citet{oskinova+10} convention: ON2 is the whole star-forming complex sized about a quarter degree, whose northern and southern halfs are respectively named ON 2N and ON 2S; the brightest (in the mid-infrared) H\textsc{ii} region in ON 2S, sized $\sim 2'$, is G75.77+0.34; and the $\sim 10''$-sized compact H\textsc{ii} region at the northeastern tip of G75.77+0.34 is named Cygnus 2N.} is a star-forming complex whose southern half, ON 2S, is projected close to the Berkeley 87 center. The main components of ON 2S are the G75.77+0.34 H\textsc{ii} region and the Cygnus 2N site of massive star formation. The later hosts several water masers whose distances have been measured by \citet{ando+11,xu+13,moscadelli+19} through trigonometric parallaxes, yielding $(3.83 \pm 0.13)~\mathrm{kpc}$, $3.56^{+0.49}_{-0.38}~\mathrm{kpc}$, and $(3.72 \pm 0.43)~\mathrm{kpc}$, respectively.

[DB2001] Cl05 is an infrared star cluster that seems to be embedded in the G75.77+0.34 cloud. The overdensity of reddened stars was discovered independently by two teams, \citet{dutra-bica01} and \citet{comeron-torra01}. While the former simply claimed that this cluster is ``related to OH maser ON2, background of cluster Be87'' without further explanation, the latter discussed ON2, Berkeley 87 and [DB2001] Cl05 as part of the same complex. No additional research on the embedded cluster was published until \citet{skinner+19} detected X-ray emission from several of the reddened stars. These authors assumed the same distance as the aligned water masers measured by \citet{xu+13}, $3.5~\mathrm{kpc}$.

To characterize a young cluster or star-forming region, it is common practice to measure properties like extinction and distance to one or few components, and then apply the results to the whole system. This simplification often works fine, since alignments of cluster-forming regions are infrequent, however this is not the case for the ON2 line of sight. Nevertheless, several works on Berkeley 87 and ON2 have proceeded in that way \citep{giovannelli02,bhavya+07,oskinova+10,binder-povich18} to obtain results or ellaborate discussions, which would be called into question if new distances were taken into account. Even though \citet{skinner+19} correctly pointed out the incompatible distances for Berkeley 87 and ON2, their implicit assumption that [DB2001] Cl05 is physically connected to the aligned masers is still unproven.

In this first paper, we address the line-of-sight overlap problem through a new approach that deals with large distance and extinction ranges in a consistent way. The intended outcome is a multi-wavelength census of objects potentially belonging to one of the overlapped young populations.

\section{Observations and data reduction}

This research makes use of deep imaging and photometry from various instruments whose spatial coverages (excluding all-sky surveys) are shown in Fig. \ref{fig:coverage}. Raw observations that were specifically reduced for this work (Sections \ref{sec:nir} and \ref{sec:xrays}) are complemented with fully-calibrated data that were made publicly available by other teams (Sections \ref{sec:mir} and \ref{sec:fir}). In Sect. \ref{sec:gaia}, the results are merged into a multi-wavelength point-source catalog whose astrometry is recalibrated with the Early Data Release 3 (EDR3) of \textit{Gaia} \citep{gaia+20}.

\subsection{Near-infrared}\label{sec:nir}

The aforementioned $15' \times 15'$ field, centered at $\alpha=20^\mathrm{h} 21^\mathrm{m} 45.1^\mathrm{s}; \delta=+37^\circ 21' 42''$, was observed in July 7, 2009, with the OMEGA2000 camera mounted on the 3.5-meter telescope of the Calar Alto Observatory, Spain. $J$-, $H$-, and $K$-band images were obtained with seeing FWHM values of 1.11, 1.23, and 1.15 arcsec, respectively.

The near-infrared images from the OMEGA 2000 camera were processed with a modified version of the FLAMINGOS pipelines \citep{levine06,romanzuniga06}. These pipelines are based on IRAF/Fortran and IDL scripts. A first pipeline is used to reduce the raw data through
a process that includes instrumental signal subtraction, illumination correction and two passes of sky subtraction before applying a combination of the dithered exposures
into a mosaic by means of an optimized centroid determination. Then a second pipeline provides identification and extraction of most sources present in the combined mosaics using
the SExtractor algorithm \citep{bertin-arnouts96} with a Gaussian filter and a 64 level deblending. Then, PSF and aperture photometry are performed on all the detected sources using
IRAF/DAOPHOT. An astrometric solution is provided by cross matching with 2MASS \citep{skrutskie+06} catalogs of the same observed region. For this dataset, we confirmed that the PSF photometry has significantly better quality than the aperture photometry measurements, mostly due to source crowding. Therefore, we decided to discard the aperture-based magnitudes. Finally, we used the \texttt{TOPCAT} software \citep{taylor05} for astrometric matching of the $J$-, $H$-, and $K$-band PSF catalogs (as well as for subsequent catalog matches in this work); a $0.75''$ tolerance for the sky position error was carefully chosen on the basis that higher separations mainly produced spurious coincidences in regions of high stellar density.

\begin{figure}
	\centering
	\includegraphics[width=0.39\textwidth, bb=4 4 278 702]{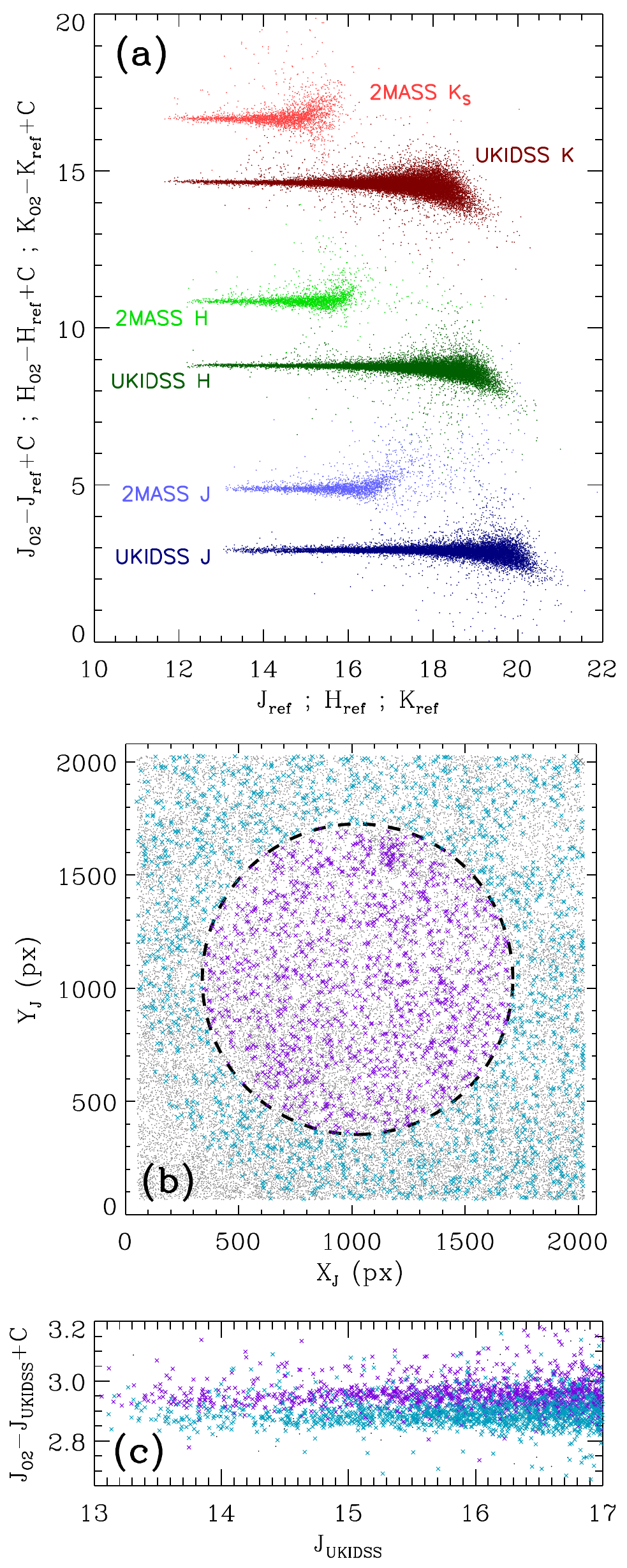}
	\caption{Photometric calibration of OMEGA2000 (``O2") data. \textit{(a)} Comparison of calibration diagrams using 2MASS or UKIDSS as the reference (``ref") survey. \textit{(b)} Location of $J$-band point sources (colored crosses: $J_\mathrm{UKIDSS} < 17$; $\sigma_J < 0.035$), relative to the flat-field artifact see Sect. \ref{sec:nir}. \textit{(c)} Close-up view of pannel \textit{a} showing the $J$-band bimodality; symbols are as in pannel \textit{b}. \label{fig:CAHAcalib}}
\end{figure}

The resulting source list was matched against the 2MASS and UKIDSS \citep{lawrence+07} catalogs for photometric calibration purposes. We performed a comparative analysis that lead us to prefer calibrating with UKIDSS \citep[i.e. WFCAM photometric system][]{Hewett+06}, for the following reasons. First, the OMEGA2000 and UKIDSS $K$ bandpasses are quite similar\footnote{Photometric filter data were retrieved from the SVO Filter Profile Service (\url{http://svo2.cab.inta-csic.es/svo/theory/fps}).}, while the 2MASS $K_S$ filter lacks the longest wavelengths ($\gtrsim 2.3~\mu m$), making equivalences more problematic. Second, UKIDSS covers a broader dynamic range, as displayed in the calibration diagrams (Fig. \ref{fig:CAHAcalib}a). Third, the lower dispersion of UKIDSS calibration allows a more accurate zero point determination. In fact, this lower dispersion makes possible the detection of a small bimodality in the J-band zero point ($\Delta J_\mathrm{ZP} \sim 0.07$; Fig. \ref{fig:CAHAcalib}c) that would be unnoticed in the 2MASS calibration. We found out that the bimodality is spatially correlated with the circular shape of a high-illumination artifact (Fig. \ref{fig:CAHAcalib}b) in the OMEGA2000 $J$-band flat-field image. We solved this issue by setting two different zero points in the appropriate regions. Finally, all the zero points in the three bands were applied after having discarded any significant color terms.

For bright sources suffering from saturation or nonlinearity problems (specifically, $J < 13.50$; $H < 12.65$; $K < 12.25$ in the OMEGA2000 field, and $J<11.5$; $H<12.04$; $K=10.5$ in the control field), 2MASS photometry was used instead. The 2MASS magnitudes were converted into the WFCAM photometric system according to the \citet{hodgkin+09} transformations.

\subsection{X-rays}\label{sec:xrays}

X-ray data of the Berkeley 87 / ON 2S region were obtained with the Advanced CCD Imaging Spectrometer (ACIS), onboard the \textit{Chandra} X-ray Observatory, in two different epochs: February 2, 2009 (ObsID: 9914; PI: Skinner) and August 13, 2016 (ObsID: 18083; PI: Skinner). The ACIS-I configuration was used, and the net exposure times were 70.15 and 69.07 ks, respectively. Results for these observations have been only published in a partial way: \citet{sokal+10, skinner+19} analized only a limited number of X-ray sources in the field, while \citet{townsley+19} published a exhaustive catalog, but for the 2009 observation alone. Therefore, we have carried out our own reduction of both epochs that leads to a homogeneous list of sources in the overlapping region (which includes the [DB2001] Cl05 cluster; see Fig. \ref{fig:coverage}). The raw data were downloaded from the \textit{Chandra} Data Archive, and processed with version 4.9 of the \textit{Chandra} Interactive Analysis of Observations (\texttt{CIAO}) software \citep{fruscione+06}, following the steps listed below.

For each dataset, we obtained a level $2$ event file with updated calibration files using the \texttt{chandra\_repro} script. Then, images in the broad (0.5 - 7 keV), soft (0.5 - 1.2 keV), medium (1.2 - 2 keV), and hard (2-7 keV) energy bands were extracted for binning factors of 1, 2, and 4 pixels through the \texttt{fluximage} tool. The point-spread function, which experiences severe variations along the ACIS-I field of view due to geometric distortion, was computed by \texttt{mkpsfmap} for each broad-band image, choosing standard values for enclosed count fraction, 0.393, and monochromatic energy, 1.49 keV. The next step consisted of detecting sources on the broad-band images through the \texttt{wavdetect} algorithm \citep{freeman+02}, taking wavelet scales of 1, 2, 4, 8 and 16 pixels.

\begin{figure}
	\centering
	\includegraphics[width=0.44\textwidth]{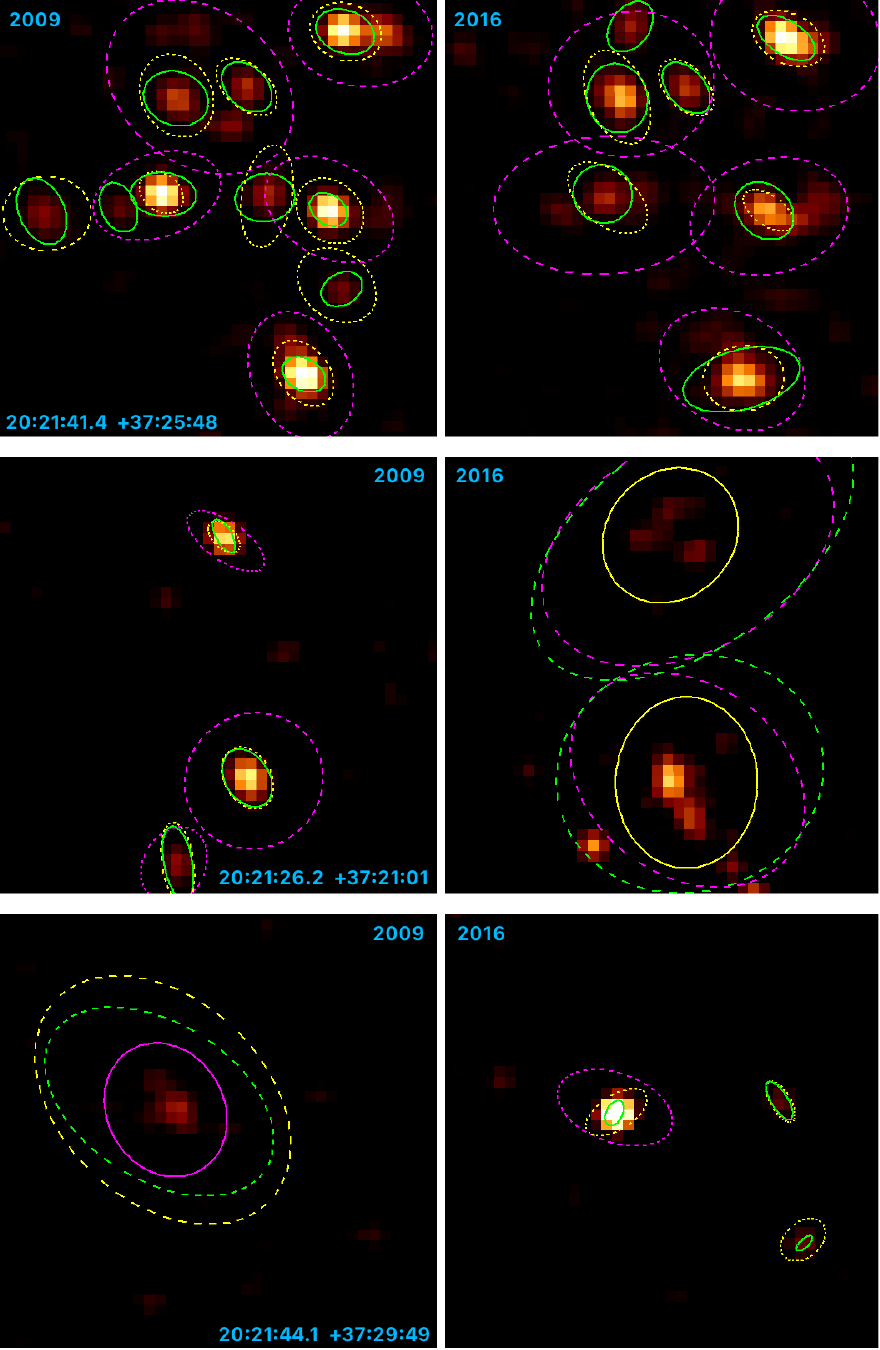}
	\caption{Comparison of X-ray detection ellipses in three $40'' \times 40''$ closeups, centered at the coordinates that are printed in blue, of the 2009 (left) and 2016 (right) \textit{Chandra}/ACIS fields, using binning factors of 1 (green), 2 (yellow), and 4 (purple) pixels. Ellipses that were selected for the final source list are drawn with solid lines. The background images are 2-pixel binned. The \textit{upper} pannels show a portion of the [DB2001] Cl05 cluster; the middle pannels, a region near the edge of the 2016 field; and the lower pannels, a region near the 2009 edge. \label{fig:Xred}}
\end{figure}

The use of 3 different binning factors is intended for optimization of astrometric accuracy, which is crucial to counterpart identification in such a crowded field. The optimal binning factor for source extraction was decided on a case-by-case basis, as illustrated in Fig. \ref{fig:Xred}. In general, the 1-pixel binning performs better in crowded regions or bright sources, and higher binning factors are preferrable near the edges of the ACIS-I field (which are severely affected by geometric distortion).

Once sources were extracted for the 2009 and 2016 fields, the two respective source lists were spatially matched, allowing for a maximum uncertainty of $4''$. This value was chosen as a compromise between source separations in the densest regions (e.g. in the upper pannels of Fig. \ref{fig:Xred}), and the position errors of \textit{Chandra} catalog sources that were measured by \citet[][see their Fig. 4]{rots-budavari11} for large off-axis angles. In the process, we detected that the 2016 field was shifted by $\Delta \alpha \approx 0.46'' ; \Delta \delta \approx 0.30''$, relative to the 2009 epoch, which is in turn better aligned with the OMEGA2000 astrometry. This shift was corrected so that both epochs are spatially matched.

The final source list contains a total of 247 point sources within the $15' \times 15'$ OMEGA2000 field, 54 of which are detected in both epochs. For these common sources, we assigned the coordinates of the smallest detection ellipse.

We note that the small number of two-epoch detections should not be interpreted in terms of variability. On the one hand, the two ACIS fields overlap only in part: the solid angle in common covers 45 \% of our near-infrared field of view (see Fig. \ref{fig:coverage}). On the other hand, detection depends strongly on geometric distortion, which can change dramatically between epochs for the same object, as explained above and shown in Fig. \ref{fig:Xred} (especially in the middle and lower pannels). Unfortunately, distinguishing between actual variability and distortion as the cause of a non-detection would require an analysis of upper limits that is beyond the scope of this paper.

Determination of X-ray fluxes is postponed to the second paper of this series, where comprehensive results for extinction from infrared counterparts will allow to obtain intrinsic fluxes. 

\subsection{Mid infrared} \label{sec:mir}

Mid-infrared images and photometric catalog were downloaded from the Cygnus-X \textit{Spitzer} Legacy Survey project website\footnote{\url{http://irsa.ipac.caltech.edu/data/SPITZER/Cygnus-X/}; see also the data delivery document at \url{http://irsa.ipac.caltech.edu/data/SPITZER/Cygnus-X/docs/CygnusDataDelivery1.pdf}.}. The survey catalog includes the 3.6, 4.5, 5.8 and 8.0~$\mu \mathrm{m}$ IRAC bands, and the $24~\mu \textrm{m}$ MIPS band.

Since Berkeley 87 is located near a corner of the surveyed region, the spatial coverage of our $15' \times 15'$ field of interest is partial in some \textit{Spitzer} bands. Specifically, the southwestern corner of the OMEGA2000 field was excluded from the 4.5 and 8.0~$\mu \mathrm{m}$ observations (see Fig. \ref{fig:coverage}). We dismiss the MIPS observations as they only cover a small portion of the near-infrared field, around its northeastern corner. 

\subsection{Far infrared} \label{sec:fir}

The ON2 region was observed by the \textit{Herschel} Space Observatory \citep{pilbratt+10} in April 11, 2012 (observation IDs: 1342244190 and 1342244191; PI: Molinari) and April 12, 2012 (observation IDs: 1342244166 and 1342244167; PI: Molinari). These observations were carried out in Parallel Mode, producing simultaneous scans at the $70~\mu \mathrm{m}$ and $160~\mu \mathrm{m}$ bands of the PACS camera \citep{poglitsch+10}, and the $250~\mu \mathrm{m}$, $350~\mu \mathrm{m}$, and $500~\mu \mathrm{m}$ bands of the SPIRE instrument \citep{griffin+10}. Due to the $\sim 21'$ shift between the PACS and SPIRE pointings in Parallel Mode, the spatial coverage was complete only for the April 11 SPIRE map and the April 12 PACS image, therefore only this combination of observation is employed. The four level-2.5 calibrated maps (one for each instrument and each date) and the corresponding point-source catalogs (only available for the PACS bands) were downloaded from the \textit{Herschel} Science Archive\footnote{\url{http://archives.esac.esa.int/hsa/whsa/}}.

\subsection{Multi-wavelength merging and astrometric recalibration} \label{sec:gaia}


In a first attempt of spatial matching between OMEGA2000 and \textit{Spitzer} sources, we detected an enhanced dispersion in the R.A. direction, together with an average offset of $0.38''$ to the West. After correcting the offset, we matched both catalogs within a box, $\sigma_\alpha = \pm 1.2''$; $\sigma_\delta = \pm 0.75''$, that takes into account the anisotropic dispersion. Because of the better resolution of our near-infrared data, the OMEGA2000 coordinates were chosen for the objects in common.

Once the near- and mid-infrared catalogs were joined, the \textit{Gaia} EDR3 catalog was downloaded from the \textit{Gaia} archive\footnote{\url{https://archives.esac.esa.int/gaia}}. \textit{Gaia} counterparts for our infrared point sources were searched in a radius of $0.75''$, consistently with previous position matches. The median values $\alpha_\mathrm{IR} - \alpha_\mathrm{Gaia} = 0.103''$ and $\delta_\mathrm{IR} - \delta_\mathrm{Gaia} = 0.162''$ were used for the recalibration; the respective standard deviations, 54 mas and 76 mas, prove the excellent performance of the OMEGA2000 relative astrometry.


\begin{figure}
	\centering
	\includegraphics[width=0.44\textwidth, bb=21 7 324 266]{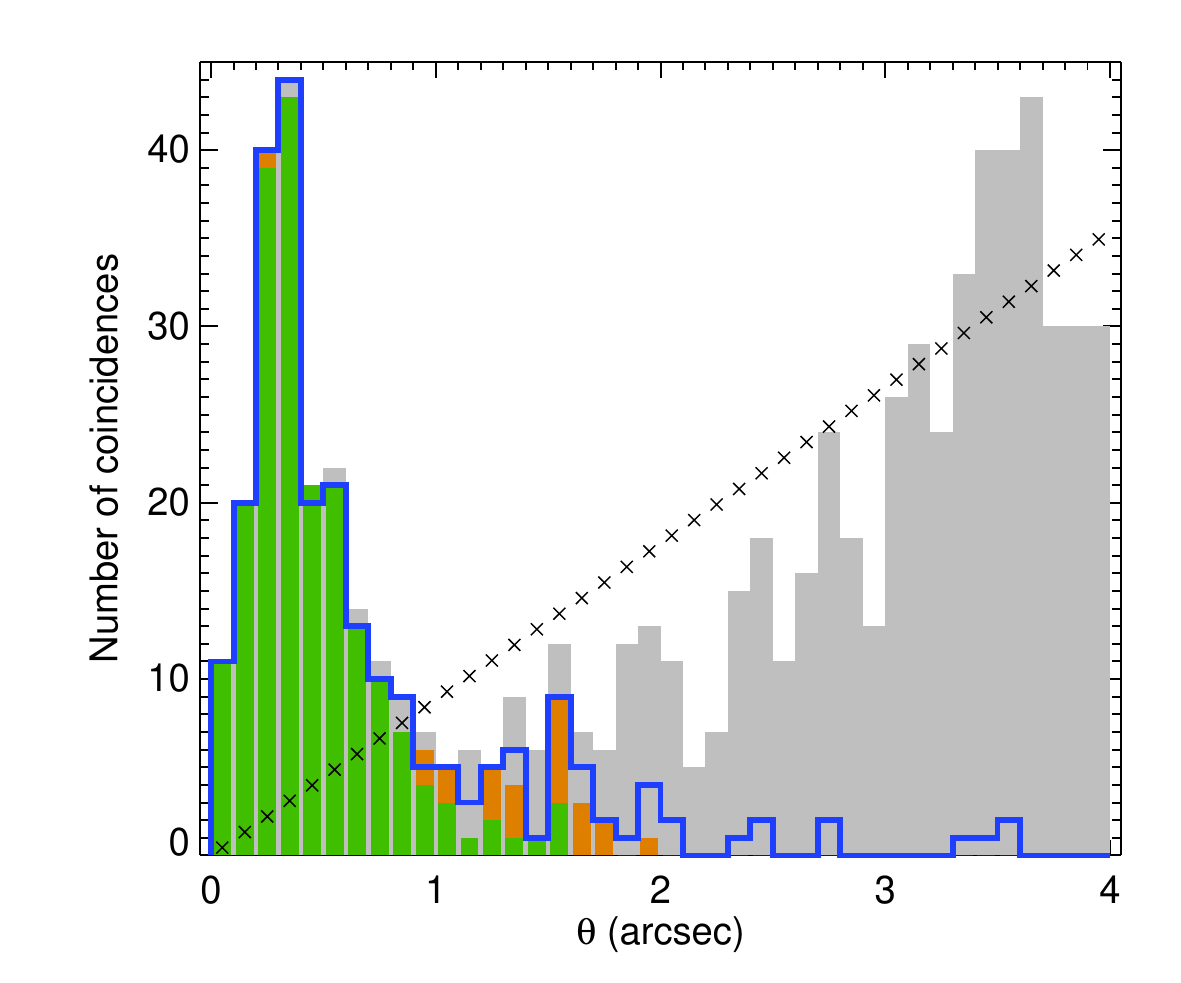}
	\caption{Histogram of angular separations between X-ray and infrared point sources (after absolute astrometric calibration with \textit{Gaia}). Gray bars include all coincidences, and the blue line only accounts for the nearest infrared neighbor of each X-ray source. Green bars are the number of infrared sources that were finally accepted as counterparts, while the orange portions were marked as doubtful. As a reference, black crosses show the case of random distribution of sources (indicative of a case where all coincidences are spurious). \label{fig:separhisto}}
\end{figure}

Merging the resulting catalog with the \textit{Chandra} and \textit{Herschel} data was not straightforward, owing to astrometric uncertainties that exceed the typical angular separations between near-infrared point sources. The \textit{Chandra} case is particularly problematic because astrometric accuracy strongly changes with position within the field of view.

Fig. \ref{fig:separhisto} shows angular distances ($\theta$) between \textit{Chandra} sources and data points from the infrared catalog. In principle, the gray histogram would lead us to interpret the minimum at $\sim 1''$ as a clear transition between genuine and spurious coincidences. However, examination of X-ray sources whose closest neighbor separations are $\theta_\mathrm{nearest} > 1''$ revealed that the longer $\theta_\mathrm{nearest}$ is, the larger the detection ellipse appears, up to a limit of $\theta_\mathrm{nearest} = 2.0''$. This entails that real counterparts are still likely to be found within that range due to variable distortion. Consequently, $\theta_\mathrm{nearest} \leq 1''$ infrared counterparts are classified as likely, provided that no additional source is present in $\theta \leq 2''$. The remaining X-ray sources fulfilling $\theta \leq 2''$ were addressed on a case-by-case basis, using the following criteria: position and size of the detection ellipse, relative to the location of the candidate counterpart; position of the secondary X-ray detection (i.e. in the alternative epoch), if any; or mid-infrared brightness. 200 infrared sources were selected as likely counterparts of \textit{Chandra} sources, and 24 were marked as doubtful. We could not determine any counterpart for the remaining 23 X-ray sources located within the $15' \times 15'$ near-infrared field.

\begin{table}
\centering
\caption{Stars of known spectral type in the OMEGA2000 field} \label{tab:sptyp}
\begin{tabular}{l l l l}
\hline \hline
TF82 ID & SIMBAD ID & Spectral type & References \\
\hline
{\it Be87-3}  & {\it HD 229059}	      & B1-2 Iabpe   & 1, 3, 4, 5	\\
{\it Be87-4}  & {\it LS II +37 73}    & B0.2 III     & 	3                \\
{\it Be87-9}  & {\it VES 203}		  & B0.5 V	     & 	3                \\
     Be87-11  & TYC 2684-25-1         & F8 V	     & 5                 \\
{\it Be87-13} & {\it TYC 2684-199-1}  & B0.5 III     & 	3                \\
{\it Be87-15} & {\it V439 Cyg}        & B1.5 Ve  	 & 	4 			    \\
{\it Be87-16} & {\it ALS 18760}		  & B2 V	 	 & 	3                \\
{\it Be87-18} & {\it ALS 18761}		  & B1 V	 	 & 	3                \\
     Be87-20  & BD+36 4031   	      & K0 III 	     & 5                 \\
     Be87-21  & Hen 3-1885	          & A0 V	 	 & 6                 \\
{\it Be87-24} & {\it ALS 18762}		  & B1 Ib	     & 	3                \\
{\it Be87-25} & {\it BD+36 4032} 	  & O8.5-9 III-V & 	4, 5 		    \\
{\it Be87-26} & {\it ALS 18763}		  & B0.5 I	     & 	3                \\
{\it Be87-29} & {\it WR 142}		  & WO2	 	     & 	5                \\
{\it Be87-31} & {\it TYC 2684-133-1}  & B1 V	 	 & 	3                \\
{\it Be87-32} & {\it TYC 2684-43-1}   & B0.5 III	 & 	3                \\
     Be87-37  & HD 229105		      & K2 II	     & 6                 \\
{\it Be87-38} & {\it VES 204}		  & B2 III	     & 	3                \\
{\it Be87-68} & {\it AS 407}		  & B2 e	 	 & 	2                \\
\hline
\end{tabular}
\tablefoot{First column numbers are identifiers from \citet{turner-forbes82}. Objects in italics are Berkeley 87 cluster members.\\
\textbf{References.} (1) \citet{majaess+08}; (2) \citet{mathew+12}; (3) \citet{massey+01}; (4) \citet{negueruela04}; (5) \citet{turner-forbes82}; (6) \citet{voroshilov+76}.}

\end{table}

Regarding \textit{Herschel}, the $70~\mu \mathrm{m}$ and $160~\mu \mathrm{m}$ catalogs were matched with maximum angular separations of $5''$. All $70~\mu \mathrm{m}$ detections were searched for counterparts in the [5.8] or [8.0] \textit{Spitzer} bands within a $4''$ radius. Since only nine \textit{Spitzer} counterparts were found, all of them were inspected. All these far- to mid-infrared matches were finally kept as positive except one \textit{Herschel} source that is compatible with the Cygnus 2N compact H\textsc{ii} region. This special case is separately discussed in Sect. \ref{sec:cyg2n} because of its relevance as a distance estimator for the ON2 region.

The final multi-wavelength list of sources contains 47090 objects in total. In order to provide cross-identifications from the literature, coordinates of previously studied objects were obtained from the SIMBAD database \citep{wenger+00} and matched with our combined source list with a $1''$ tolerance. These include 19 stars of known spectral type, which are listed in Table \ref{tab:sptyp}. Finally, identifiers from the recent X-ray catalog by \citet{skinner+19} were also added to our catalog by matching their coordinates in a $1''$ radius.

\subsubsection{The Cygnus 2N region} \label{sec:cyg2n}

\begin{figure}
	\centering
	\includegraphics[width=0.46\textwidth, bb=31 13 515 463]{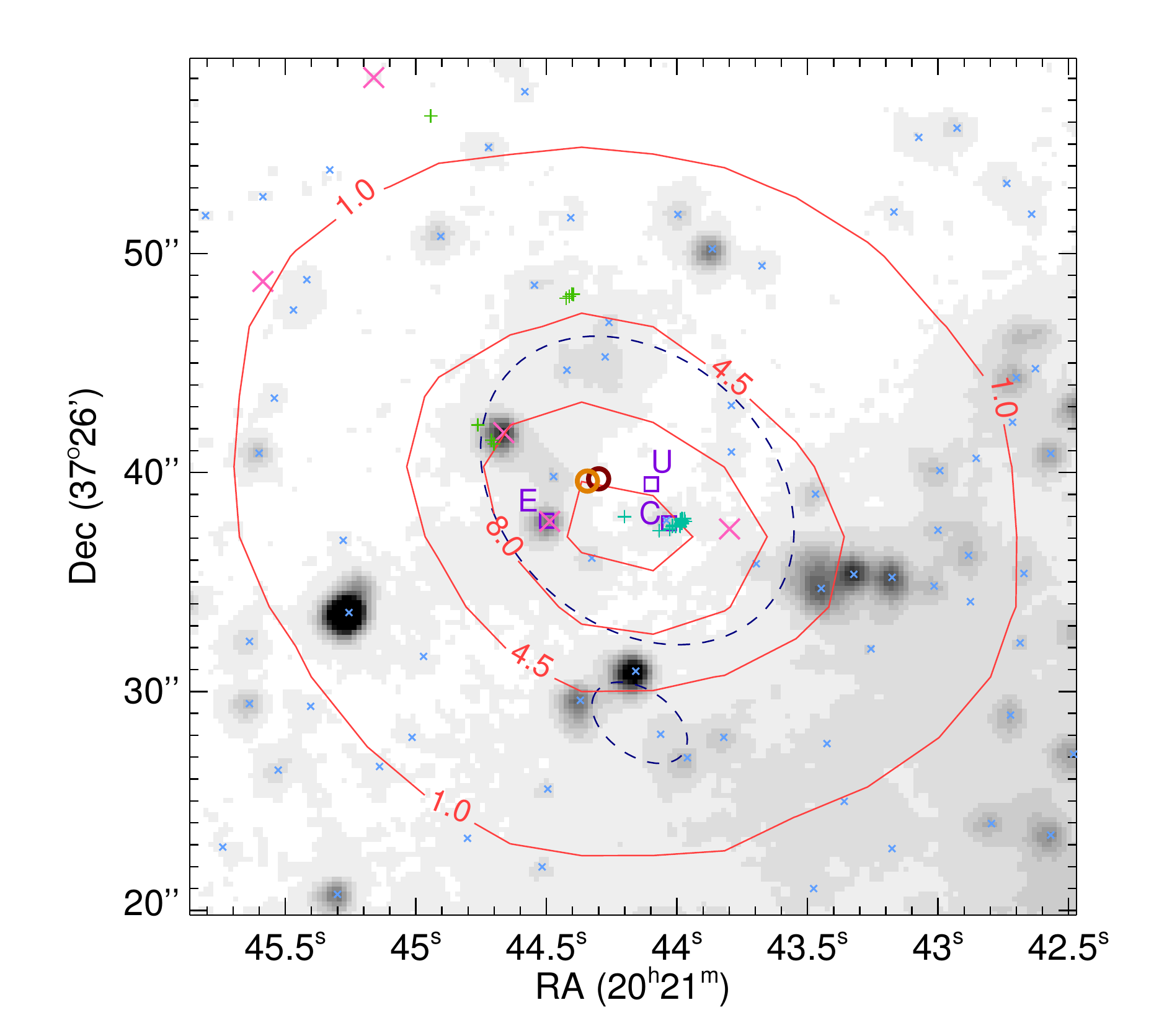}
	\caption{OMEGA2000 $K$-band image (grayscale) and point-source catalog (small blue crosses) of the Cygnus 2N region. The \textit{Herschel}/PACS $70~\mu \mathrm{m}$ map is overplotted as contours with an interval of $3.5~\mathrm{Jy}/\mathrm{arcsec}^2$; the positions of $70~\mu \mathrm{m}$ and $160~\mu \mathrm{m}$ detections according to the \textit{Herschel} public catalog are marked as orange and red circles, respectively. \textit{Chandra} detection ellipses are drawn as blue dashed lines, and infrared sources detected by \textit{Spitzer} at wavelengths larger than $5~\mu \mathrm{m}$ are shown as large pink crosses. The remaining symbols indicate objects reported by \citet{sanchezmonge+13}: green and turquoise plus signs are methanol and water masers, respectively, and purple squares are the centimeter continuum sources that were named CORE (C), EAST (E), and UCHII (U) by these authors. \label{fig:cyg2n}}
\end{figure}

Fig. \ref{fig:cyg2n} displays multi-wavelength source positions over a $K$-band image of the Cygnus 2N H\textsc{ii} region and its suroundings. The brightest PACS point source in the ON2 region ($f_{70 \mu \mathrm{m}} = 2066~\mathrm{Jy}$; $f_{160 \mu \mathrm{m}} = 2127~\mathrm{Jy}$) is seemingly coincident with the Cygnus 2N radio continuum source \citep{matthews+73,harris74} which is, in turns, subdivided in three compact sources of different evolutionary state \citep{sanchezmonge+13, cho+16}. Also, three \textit{Spitzer} sources with very red colors \citep[one of them being coincident with the EAST ultra-compact H\textsc{ii} region, as already noted by][]{sanchezmonge+13} are located very close to the far-infrared peak. All of them form a structure that is elongated in the same direction than the \textit{Herschel} source, as shown by red contours in Fig. \ref{fig:cyg2n}. Such geometry might be indicating that far-infrared radiation in this small region comes from a group of objects that are not resolved by \textit{Herschel}. In any case, the uncertainty of the $70~\mu \mathrm{m}$ detection is larger than the separations between point sources in Fig. \ref{fig:cyg2n}, as revealed by the discrepancy between the peak position and the PSF-fitted coordinates \citep[see][]{marton+17} from the PACS catalog. Likewise, the \textit{Chandra} detection in the central part of Fig. \ref{fig:cyg2n} could also be related to various objects belonging to the Cygnus 2N star-forming region, given the extended detection ellipse size encompassing multiple infrared sources.

The above presented arguments led us to consider both the  \textit{Herschel} and \textit{Chandra} detections as counterparts of the whole Cygnus 2N H\textsc{ii} region. Meanwhile, the closest mid-infrared source, which was initially matched with the \textit{Herschel} source, is now considered as a different catalog entry that matches the EAST object.

Nevertheless, a caveat about the Cygnus 2N X-ray counterpart must be added: this faint source is only detected by \texttt{wavdetect} in the 4-pixel binned image of the 2016 observation, where this source is placed at the edge of the gap between ACIS-I detectors. Hence, we cannot discard that this detection is caused by a reduction artifact. Still, coincidence with the Cygnus 2N region points that this \textit{Chandra} source is real, since X-ray emission is commonly associated with regions of massive star formation \citep{townsley+14,townsley+18}.

\section{Basic observational features and practical definitions}

\begin{figure*}
	\centering
	\includegraphics[width=0.41\textwidth]{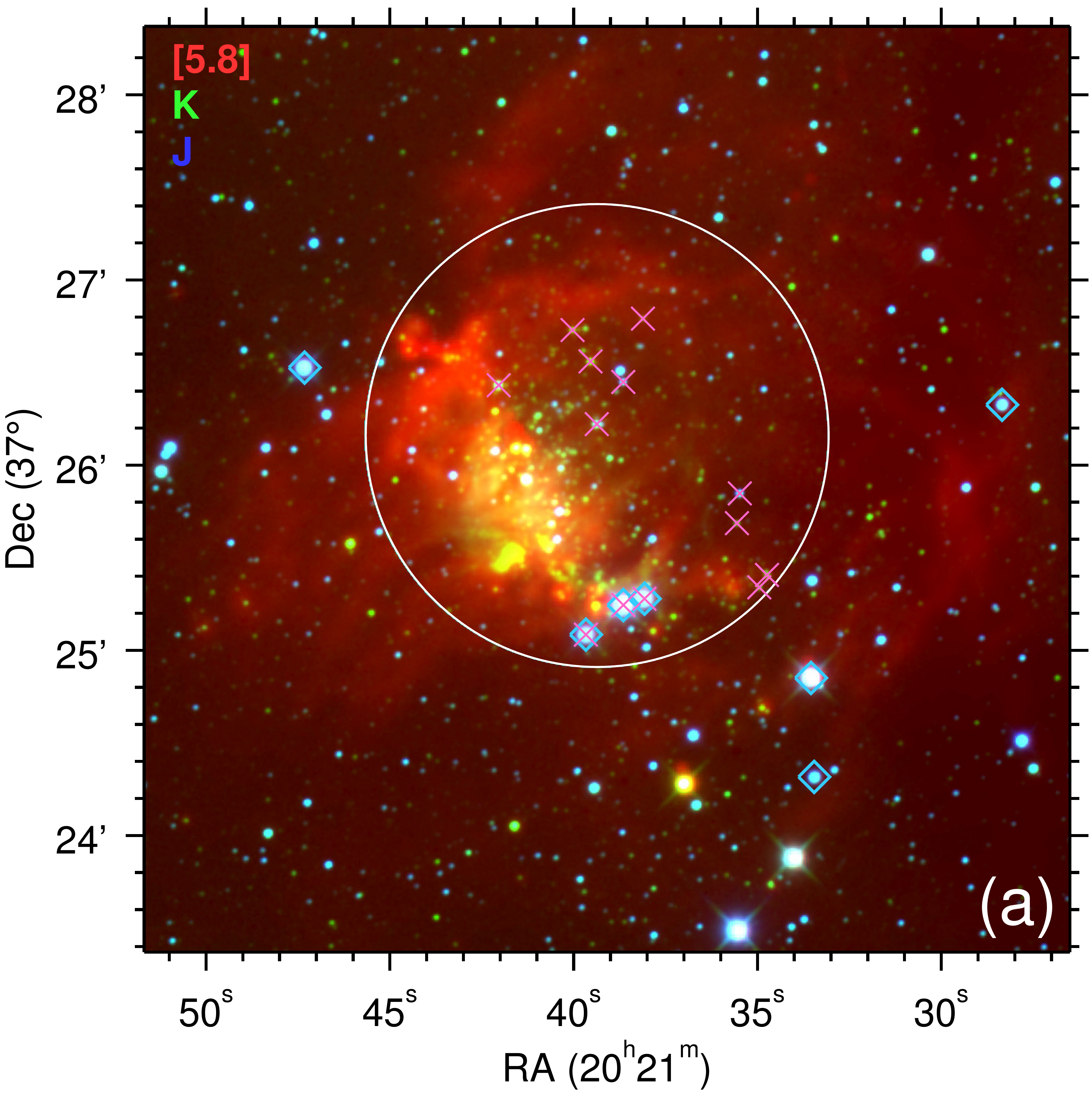}
	~
	\includegraphics[width=0.42\textwidth,bb=38 3 405 370]{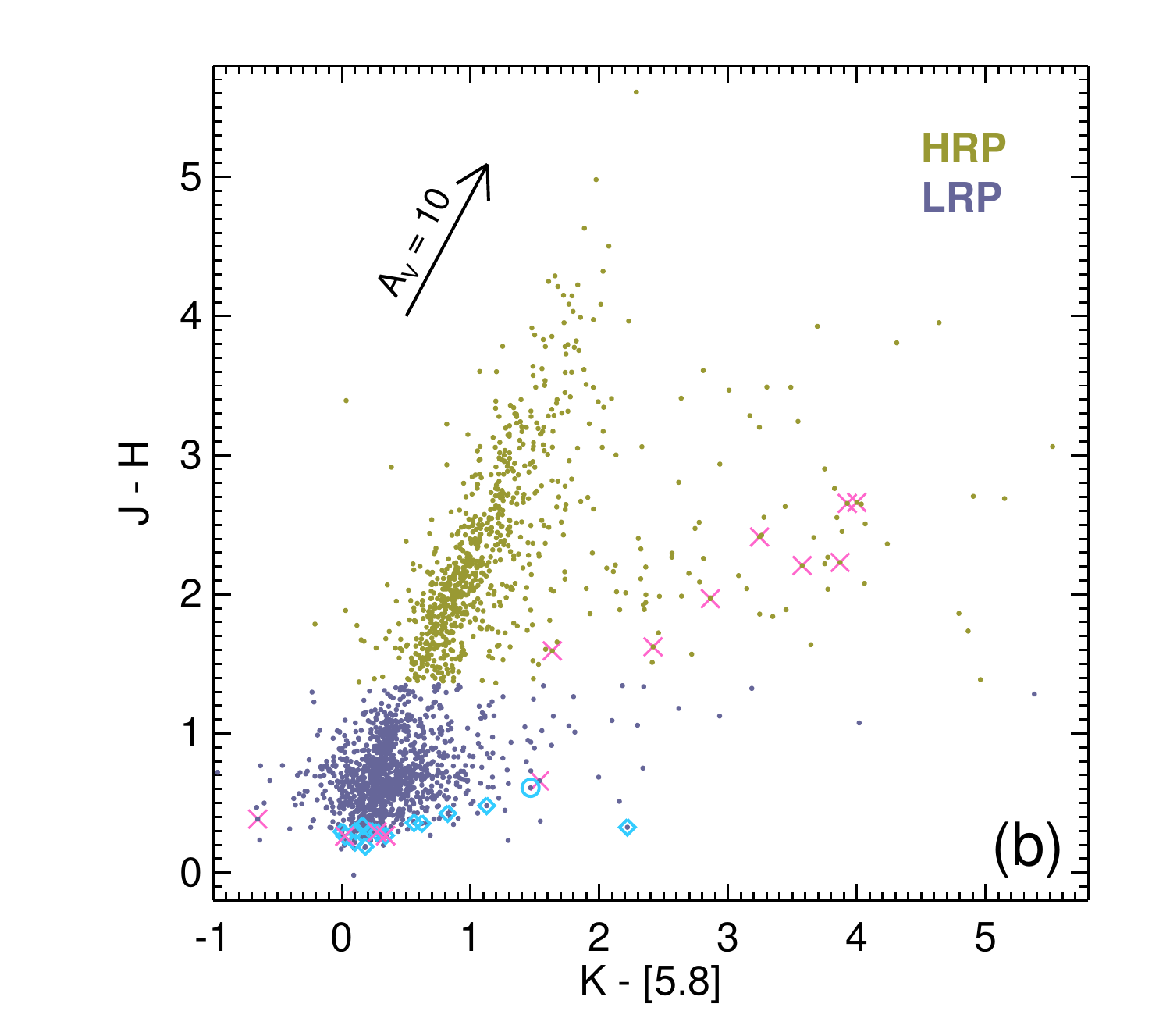}
	~
	\includegraphics[width=0.41\textwidth]{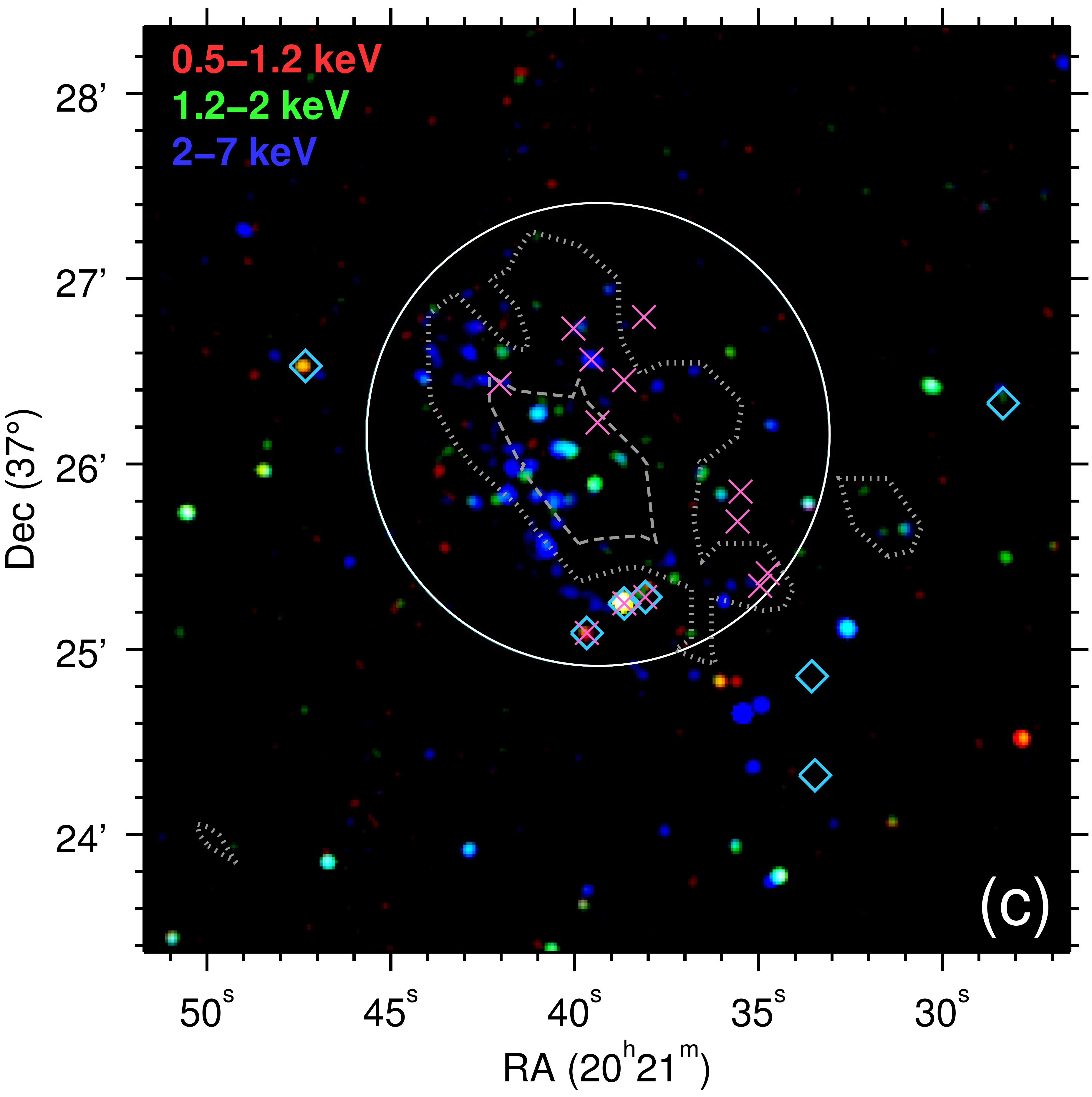}
	~
	\includegraphics[width=0.42\textwidth]{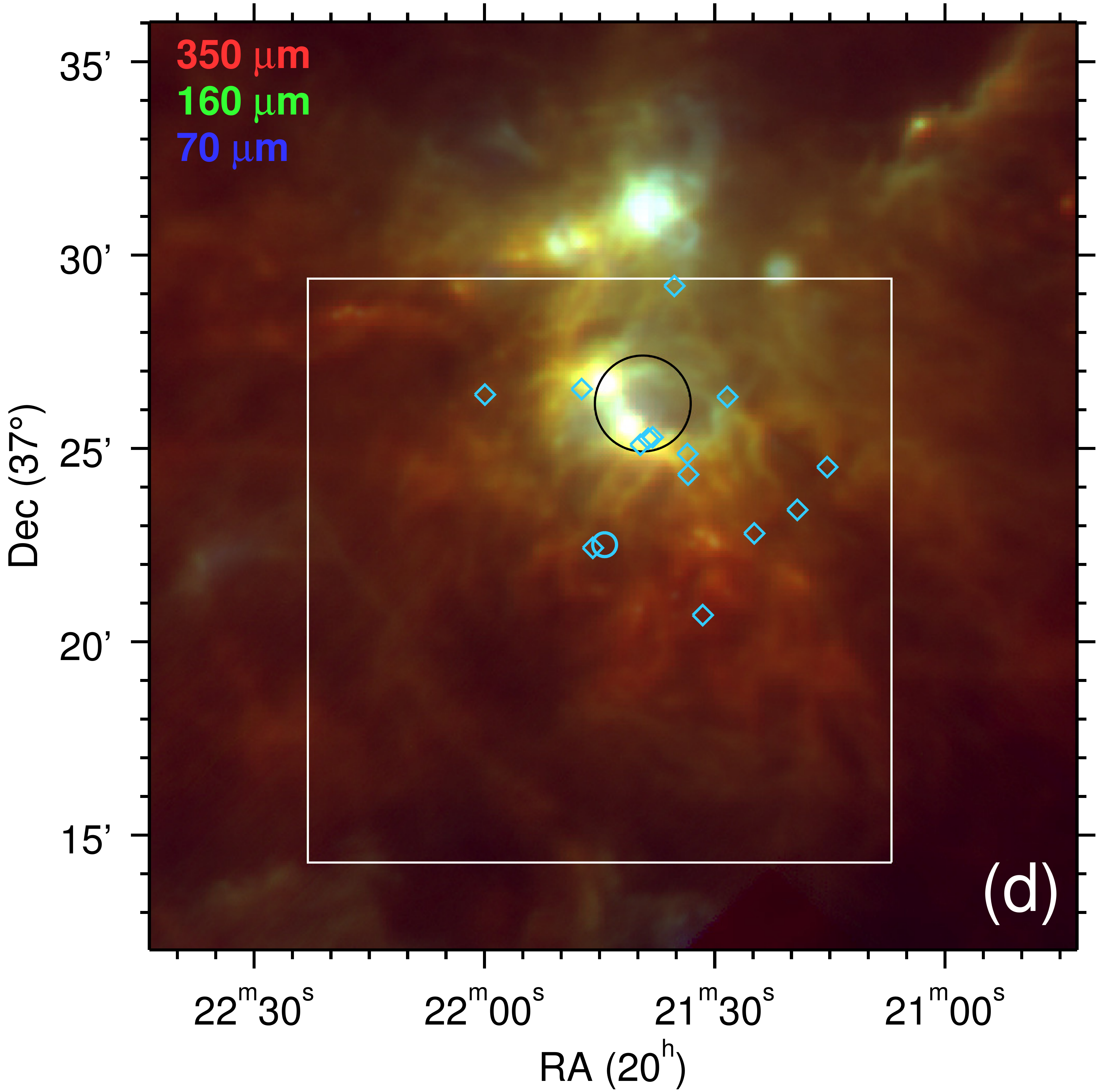}
	\caption{Multi-wavelength RGB images (colored as indicated in the upper left corner of each image) of the Berkeley 87 / ON2 region, and infrared color-color diagram of the OMEGA2000 field. In all the images, the Cl05 region is enclosed by a white or black circle. In the \textit{Herschel} RGB image \textit{(d)}, the OMEGA2000 field coverage is shown as a white square. An infrared closeup of this field, centered on [DB2001] Cl05, is displayed in pannel \textit{(a)}, and its \textit{Chandra} counterpart is shown in pannel \textit{(c)}. The latter is made of 2-pixel binned images from the 2009 observation; the density of $J-H>1.7$ sources is drawn as gray contours (dotted~$=55~\mathrm{arcmin}^{-2}$; dashed~$=110~\mathrm{arcmin}^{-2}$). In all pannels, Berkeley 87 spectroscopic members are marked as blue open diamonds, except WR 142, which is shown as a blue open circle (only apperaring in  pannels \textit{b} and \textit{d}); crosses are point sources within the Cl05 region that are simultaneously detected in $J$, $H$, $K$, and [5.8] (i.e. those that can be represented in pannel \textit{b}). \label{fig:cyg05rgb}}
\end{figure*}

In a first inspection of the multi-wavelength data (e.g. images and photometry in Fig. \ref{fig:cyg05rgb}, which will be discussed below in more detail), we find the following observational features hinting at a superposition of components. First, Berkeley 87 known members (a total of 15) seem to be among the least reddened point sources, implying that valuable information can be provided by \textit{Gaia}. Second, a conspicuous concentration of very reddened objects toward the [DB2001] Cl05 region can be only noticed at X-ray and infrared wavelengths. Based on such features, we provide below some practical definitions to isolate different populations.

\subsection{Preliminary components} \label{sec:prelim}

Isocontours enclosing sources of increasing $J-H$ colors revealed a conspicuous overdensity of very red ($J-H>1.7$) point sources around the position of the [DB2001] Cl05 cluster. As seen in Fig. \ref{fig:cyg05rgb}c, this overdensity (gray contours) is rougly coincident with the group of X-ray point sources in the G75.77+0.34 H\textsc{ii} region that was already noticed by \citet{skinner+19}. Based on this spatial distribution, we define the ``Cl05 region'' as the circle of radius 75 arcsec centered at $\alpha=305.414; \delta=37.436$ (which is drawn on all RGB images in Fig. \ref{fig:cyg05rgb}), to serve as a rough, preliminary delimitation of the embedded cluster. We remark that this definition has merely practical purposes (e.g. to compare point source properties inside and outside the overdensity) and is not aimed at anticipating any properties of the cluster.

Another striking observational feature of the field is a bimodality in the \textit{JHK} colors that is clearly seen in color-color diagrams (Fig. \ref{fig:cyg05rgb}b and others in Fig. \ref{fig:rintccd}). Since near-infrared color excesses (especially in the $J-H$ case) are mainly influenced by extinction \citep[in contrast to the mid-infrared, where intrinsic reddening becomes more important; see e.g.][]{gutermuth+08,teixeira+12}, this bimodality may be interpreted as two extinction groups approximately separated by the minimum of the distribution $J-H  \approx 1.35$, or $H-K \approx 0.65$. Hence, we define the Low-Reddening Population (LRP) as composed of $J-H \leq 1.35$ sources, or as $H-K \leq 0.65$ for those lacking J-band detection. Likewise, we define the High-Reddening population (HRP) as meeting $J-H>1.35$, or $H-K>0.65$ if $J$ is missing.

Preliminarly, Fig. \ref{fig:cyg05rgb} provides evidence that the apparent cluster pair is just a chance alignment of unrelated star formation events. The color-color diagram in pannel \textit{b} reveals a large gap along the extinction vector\footnote{The extinction vector is obtained from \citet{indebetouw+05}, and taking $A_K/A_{K_S} \approx 0.98$ (which is justified in Appendix \ref{sec:rint}).} between Berkeley 87 members (all of which belonging to the LRP), and HRP sources located within the Cl05 region. The latter show particularly large $K-[5.8]$ colors, favoring very young ages \citep[see e.g.][]{teixeira+12} for [DB2001] Cl05, in contrast to the Berkeley 87 cluster. Moreover, the distribution of hot dust (bluer colors on the \textit{Herschel} image in pannel \textit{d}) appear to be spatially compatible with the Cl05 region, and incompatible with the projected distribution of hot massive Berkeley 87 members despite the well-known feedback power of such kind of stars. The case of WR 142 is especially meaningful, as this extremely hot star and its mighty wind \citep{tramper+15, sander+19} show no heating effects on the aligned dust that is detected by \textit{Herschel}; this lack of hot matter suggests a distance discrepancy between Berkeley 87 and ON2.

\subsection{Intrinsic reddening} \label{sec:intred}

Over the course of this paper, we need to distinguish between intrinsic and interstellar color excess, for two main reasons. First, breaking this degeneracy is crucial to correctly estimate interstellar extinction towards each star, especially when only photometric information is available. Second, we aim at finding populations of Young Stellar Objects (YSOs), which are commonly revealed as intrinsically red sources. Although other photometric methods for YSO discovery and classification, based on color excess, have been already developed \citep[e.g.][]{lada+06,gutermuth+09}, such methods require an extensive wavelength coverage in the mid-infrared that is not fulfilled in the Cl05 region. For example, few \textit{JHK} sources are also detected in [5.8] within this region (pink crosses in Fig. \ref{fig:cyg05rgb}). Therefore, the incompleteness of near-infrared data demands a more flexible treatment.

Since reddening is dominated by interstellar extinction towards shorter infrared wavelengths \citep[see e.g.][]{indebetouw+05}, color-color diagrams that combine one long-wavelength color $X=X_2-X_1$ and one short-wavelength color $Y=Y_2-Y_1$ (e.g. Fig. \ref{fig:cyg05rgb}b) are useful for determining the origin of infrared color excess. We specifically refer to $(X,Y)$ diagrams where the $X$ color encompasses K-band wavelengths and longer, and the $Y$ color is part of the \textit{JHK} range. As shown by \citet{teixeira+12}, such diagrams place most of stars on the main sequence and its interstellar reddening band, while intrinsically red sources are clearly shifted to redder $X$ colors. Based on this idea, we propose to measure intrinsic reddening of each object as a weighted average of the $X$ excesses that are measured in the $(X,Y)$ diagrams where the source is present. Thus, we define the intrinsic reddening index, $R_\mathrm{int}$, as:

\begin{equation}
 \log R_\mathrm{int} = 0.1~\frac{\sum_{X,Y} W_{X,Y} S_{X,Y}}{\sum_{X,Y} W_{X,Y}}\,,
 \label{eq:rint}
\end{equation}
where each $S_{X,Y}$ is the relative shift, along the $X$ direction, from the reddening band boundary of each diagram; and $W_{X,Y}$ is the respective weight. The complete definitions of $S_{X,Y}$, $W_{X,Y}$, and the reddening band boundaries, are presented in Appendix \ref{sec:rint}, together with a discussion on which $X$ and $Y$ colors are included (or excluded). The 6 resulting color-color diagrams that are finally involved in the $R_\mathrm{int}$ calculation are displayed in Fig. \ref{fig:rintccd}.

\section{Extinction and distance} \label{sec:extinction}

This section is aimed at measuring extinction and distance for as many objects as possible in our merged photometric catalog.

Due to the unusually complex nature of the field, with multiple stellar populations located at different distances, several methods of extinction estimation are combined. On the one hand, we employ intrinsic colors of relatively nearby stars when such information can be obtained, either directly or indirectly, from the literature. On the other hand, we estimate extinction for a much larger sample by applying the \citet{majewski+11} method to our photometric data.

\subsection{Extinction from individual estimates of intrinsic color} \label{sec:individual}

First of all, we address extinction estimation for those stars whose intrinsic colors can be inferred directly from their known spectral types. All of the 19 stars in the catalog with previously published spectra are saturated in the OMEGA2000 images, therefore 2MASS photometry is used instead. Also, we take advantage of the V-band photometry published by \citet{turner-forbes82} that include all these 19 objects.

In general, we rely on the intrinsic colors published by \citet{ducati+01} for normal stars later than O9 I or B0 V. This is not applicable to the following exceptions: the O8.5-9 III-V star BD+36$^\circ$4032 for which we use \citet{martins-plez06}, and the special cases of WR 142 and V439 Cyg that will be addressed separately. Since the two cited works employ the Johnson-Glass photometric system for the near-infrared \citep{bessell-brett88}, intrinsic colors are converted into the 2MASS system through the \citet{carpenter01} transformations. For stars earlier than F, we utilize $V-K_S$ to obtain color excess ($E_{V-K_S}$), as its relative uncertainty is the lowest, thanks to the extended wavelength range and the fact that $A_{K_S} \ll A_\mathrm{V}$ (therefore making the result less dependent from the chosen extinction law). For late-type stars, computing $E_{V-K_S}$ is no longer appropriate, since the uncertainties on intrinsic colors \citep[see e.g. $V-K$ variations between adjacent spectral subtypes in][] {ducati+01} become comparable to the resulting color excess values; in these cases, $J-K_S$ is used instead.

Color excess for WR 142 and V439 Cyg cannot be accurately determined in the same way. In the former case, because of the lack of knowledge about intrinsic colors for WO stars, which are extremely rare. In the latter case, the strong photometric and spectroscopic variability of this object \citep{polcaro+89, polcaro-norci98} would make necessary simultaneous photometry and spectroscopy. Then, we prefer to adopt the $E_{B-V}(\mathrm{V439\>Cyg}) = 1.53$ value that \citet{turner-forbes82} obtained from color excesses of nearby Berkeley 87 members. Likewise, we assign WR 142 the color excess $E_{V-K_S} = 3.30$ that we computed for an angularly close cluster member, TYC 2684-133-1.

\citet{turner-forbes82} estimated $E_{B-V}$ for 19 additional objects of unknown spectra, by assuming they are B-type stars in the Berkeley 87 main sequence. This assumption is, in turn, based on the UBV color-color diagram published by these authors. We use these color excess data as well, notwithstanding that some of them would be excluded later if their \textit{Gaia} parallaxes are incompatible with the cluster.

Once color excesses for all the above adressed objects (38 in total) are determined, their extinction values are calculated as follows. First, $A_\mathrm{V}$ is obtained for the 18 stars with $E_{V-K_S}$ and $E_{J-K_S}$ values through the \citet{rieke-lebofsky85} extinction law \footnote{Again, the correction $A_K/A_{K_S} \approx 0.98$ from Appendix \ref{sec:rint} is applied, yielding $A_{K_S}/A_\mathrm{V} \approx 0.114$.\label{foot:rl85cor}}. Among these 18 objects, 13 correspond to confirmed B stars, being the only secured members whose color excesses have been determined independently (unlike WR 142 or V439 Cyg, based on nearby stars). Therefore, we take these 13 extinction results (all within the range $3.75 \le A_\mathrm{V} \le 5.57$, with an average of $\bar A_\mathrm{V} = 4.7$) as representative of Berkeley 87. By dividing each $A_\mathrm{V}$ results by the corresponding $E_{B-V}$ from \citet{turner-forbes82}, we obtain $\bar R_V = 2.8$ and $\sigma_{R_V} = 0.11$ for the sample of 13 confirmed B-type members. Finally, this new $R_V$ calibration is used to convert the remaining $E_{B-V}$ estimates (i. e. from V439 Cyg and the \citet{turner-forbes82} B-type candidates) into their visual extinction values.

\subsection{Extinction from the RJCE method} \label{sec:rjce}

Owing to the low number of extinction determinations so far, and the fact that nearly all of them are candidate Berkeley 87 members, a substantial extension of the extinction sample is required to separate components along the line of sight. For this purpose, we use the Rayleigh-Jeans Color Excess (RJCE) method, designed by \citet{majewski+11}. These authors provided an equation to compute $A_{K_S}$ through $H-[4.5]$, on the basis that the intrinsic value of this color, $(H-[4.5])_\mathrm{int}$, is virtually the same for the majority of spectral types, with notable exceptions (e.g. KM dwarfs) that are discussed below. Since \citet{majewski+11} employed the 2MASS H-band filter, we have adapted their equation to the OMEGA2000 filter through the appropriate calibration (Sect. \ref{sec:nir}):

\begin{equation}
 A_{K_S} = 0.918\,(H_\mathrm{OMEGA2000}-[4.5]-0.13)\,,
\end{equation}
which is divided by 0.114 (see footnote \ref{foot:rl85cor}) to obtain $A_\mathrm{V}$. We should be careful, however, not to apply the RJCE method for objects whose $(H-[4.5])_\mathrm{int}$ may deviate significantly from the nominal value of the method (0.08 when $H_\mathrm{2MASS}$ is employed; 0.13 according to our adapted version). The trivial case is the population that shows a significant intrinsic color excess in the mid-infrared, i.e. precisely what $R_{int}$ is able to distinguish. Hence, we only apply the RJCE method to objects fulfilling $R_\mathrm{int} \leq 1.25$.

According to \citet{majewski+11}, other objects with deviating $(H-[4.5])_\mathrm{int}$ values are OB stars, which are bluer, and KM dwarfs, which are redder, therefore leading to $A_{K_s}^\mathrm{(RJCE)}$ underestimations and overestimations, respectively. Such cases can be excluded by isolating the stellar populations where these objects are detectable and their extinction values can be measured by the RJCE method, as discussed below.

As OB stars are rare outside massive or intermediate-mass young ($\lesssim 10^8~\mathrm{yr}$) clusters or associations, and given that virtually no HRP sources in the Cl05 region have their extinction measurable by the RJCE method (with the $R_\mathrm{int} \leq 1.25$ constraint)\footnote{this lack of RJCE results is caused by a mixture of photometric incompleteness effects, like those described in Sect. \ref{sec:intred} and Appendix \ref{sec:photbias}, and the fact that [DB2001] Cl05 appear to be dominated by $R_\mathrm{int} > 1.25$ objects.}, we can concentrate our efforts in Berkeley 87 and its surroundings. Since the largest color excess of a cluster member, $E_{B-V} = 1.9$ \citep{turner-forbes82}, corresponds to $A_{K_S} = 0.65$ \citep[by applying][]{rieke-lebofsky85}, we can safely assume that any underestimated $A_{K_S}^\mathrm{(RJCE)}$ values for OB stars would be lower than 0.65 in any case. Consequently, we have decided to discard all $A_{K_S}^\mathrm{(RJCE)}$ results below 0.65, as some of them may be underestimated due to their unidentified OB nature. Since the photometric uncertainty of the RJCE method, $\sigma_{K_S}^\mathrm{(RJCE)}$, is significant in some cases, we only keep results that fulfill:

\begin{equation}
 A_{K_S}^\mathrm{(RJCE)} - \sigma_{K_S}^\mathrm{(RJCE)} > 0.65
 \label{eq:rjcelimit}
\end{equation}

Regarding KM main-sequence stars, these low-luminosity objects are only detectable if they are much closer than Berkeley 87, and far less extinguished. This implies that Equation \ref{eq:rjcelimit} would only be fulfilled by KM dwarfs whose $(H-[4.5])_\mathrm{int}$ color is too large to cause such an overestimate, in which case they would fail to meet the $R_\mathrm{int} \leq 1.25$ condition. The validity of this claim will be evaluated in Sect. \ref{sec:distance}, taking advantage of the \textit{Gaia} capabilities for measuring nearby, unextinguished faint sources.

As a side effect, Equation \ref{eq:rjcelimit} excludes not only OB stars and KM dwarfs, but also any other sources that are little affected by extinction. Fortunately, such objects are easily detected by \textit{Gaia}, which potentially provides data that are more useful for membership determination than extinction. 

\subsection{Gaia distances} \label{sec:Be87}

Because of the nature of \textit{Gaia} observations, parallax data are necessarily focused on populations that are observed under low or mild extinction conditions, e.g. Berkeley 87. For the purpose of separating populations along the line of sight, this allows \textit{Gaia} parallaxes to play a role that is complementary to extinction data from Sect. \ref{sec:rjce}.

\subsubsection{Parallax bias correction} \label{sec:parlxbias}

Before proceeding with distance estimation, we clarify how we deal with the systematics of \textit{Gaia} EDR3 parallaxes. \citet{lindegren+20} provide a recipe to correct the parallax bias through an approximate model, which consists of two separate functions for the five- and six-parameter (hereafter 6-p) astrometric solutions. This model is only valid for certain magnitude and color ranges; only 1830 out of 3280 sources with available EDR3 parallaxes in the OMEGA2000 field fall within these ranges.

First, we perform the bias correction for these 1830 objects through \texttt{gaiadr3\_zeropoint}\footnote{The \texttt{gaiadr3\_zeropoint} package was downloaded from \url{https://gitlab.com/icc-ub/public/gaiadr3_zeropoint}}, a Python3 implementation of the \citet{lindegren+20} recipe. The resulting bias estimates, $\varpi-\varpi^\mathrm{corr}$ (where $\varpi^\mathrm{corr}$ is the bias-corrected parallax), are always negative, with average and median values $-34$ and $-36~\mu \mathrm{as}$, respectively \citep[c.f the global EDR3 values, $-17$ and $\-21~\mu \mathrm{as}$;][]{lindegren+20}.

This evidence of assymetric bias suggests that at least a rough bias correction should be also applied to the remaining 1450 objects before estimating their distances. All these objects falling outside the validity ranges of the \citet{lindegren+20} model correspond to 6-p astrometric solutions. Therefore, the average of the above calculated bias estimates for the 6-p subset, $-37~\mu \mathrm{as}$, is adopted as a zero-order correction and applied to the aforementioned 1450 parallaxes.

\subsubsection{Procedure for distance estimation} \label{sec:procedure}

Following the recommendations of \citet{astraatmadja-bailerjones16} and \citet{luri+18}, we choose a Bayesian approach to produce distance estimates, using the Exponentially Decreasing Space Density (EDSD) prior \citep{bailerjones15}. This prior employs a single adjustable parameter, the scale length $L$, and the mode of its probability distribution is equal to $2L$. Since we aim at finding out where each object is located relative to Berkeley 87, we opt to use a single value for $L$ that makes the mode of the prior equal to a first estimate of the cluster distance, obtained through parallax inversion:


\begin{equation}
 2L = r_\mathrm{Be87}^{(0)} = 1 / \varpi^\mathrm{corr}_\mathrm{Be87}\,,
 \label{eq:scalelength}
\end{equation}

where $\varpi^\mathrm{corr}_\mathrm{Be87}$ is computed as the median of the 13 spectroscopically confirmed members with fractional uncertainties $f_\varpi = \sigma_\varpi/\varpi < 0.1$. Higher $f_\varpi$ values are avoided here since the parallax inversion becomes strongly biased as a distance estimator \citep{luri+18,bailerjones+18}. This yields $\varpi^\mathrm{corr}_\mathrm{Be87} = 0.5976~\mathrm{mas}$ and $r_\mathrm{Be87}^{(0)} = 1673~\mathrm{pc}$.

Hence, we set $L=836.5~\mathrm{pc}$ in the \texttt{TOPCAT} implementation of the Bayesian estimator (which make use of the EDSD prior), to obtain distance estimates ($r_\mathrm{est}$) from all the parallax measurements. Likewise, we compute the 15.87\% and 84.13\% quantiles ($r_{16}$, $r_{84}$) of the probability density function of the posterior \footnote{In this way, the confidence interval defined as $[r_{16}, r_{84}]$ is comparable, in terms of likelihood, to that utilized by \citet{bailerjones+18}. In both cases, the enclosed probability is 68.27\%, the same as for a Gaussian distribution within a $\pm 1\sigma$ interval centered at its maximum.}. The distance results for the 13 aforementioned Berkeley 87 members have a median of $r_\mathrm{Be87} = (1673 \pm 17)~\mathrm{pc}$.

We have to avoid the regime of high $f_\varpi$ values where the posterior is significantly influenced by the prior choice, as these would yield distance results that are spuriously consistent with Berkeley 87. The transition from the data-dominated to the prior-dominated posterior occurs at $f_\varpi \sim$~0.3 - 0.4 for the EDSD prior \citep{bailerjones15, astraatmadja-bailerjones16}. Therefore, a simple solution would be setting an $f_\varpi$ threshold well below that transition, e.g. $f_\varpi \leq 0.2$. We should make sure, however, that such a constraint would not disregard any valuable results, by checking the produced confidence intervals, $\Delta r = r_{84}-r_{16}$. In our catalog, data points fulfilling $f \approx 0.2$ yield relative uncertainties $\Delta r / r_\mathrm{est} \approx 0.5$, and we consider that any looser distance determinations would not be useful to our goals. Likewise, negative parallaxes \citep[see][]{luri+18} can be safely discarded in our case, since they always lead to $\Delta r / r_\mathrm{est} > 0.5$.

Consequently, we decide to calculate distances only for parallaxes fulfilling $0 < f_\varpi \leq 0.2$.

Finally, we must add a caveat about the impact of our approach on the validity of distance results. The above described procedure is tailored to the scientific goals of this series of papers; in particular, distances are intended to be evaluated relative to the optically visible cluster (i.e. Berkeley 87). Conversely, if optimal calculations for absolute distances were required for other purposes, a more careful election of the scale length should be made. Nevertheless, our prior choice (Equation \ref{eq:scalelength}) is expected to have a minor effect on distance results, provided that the permitted $f_\varpi$ range is well inside the data-dominated regime.

\subsubsection{Distance results vs. extinction} \label{sec:distance}

\begin{figure}
	\centering
	\includegraphics[width=0.48\textwidth]{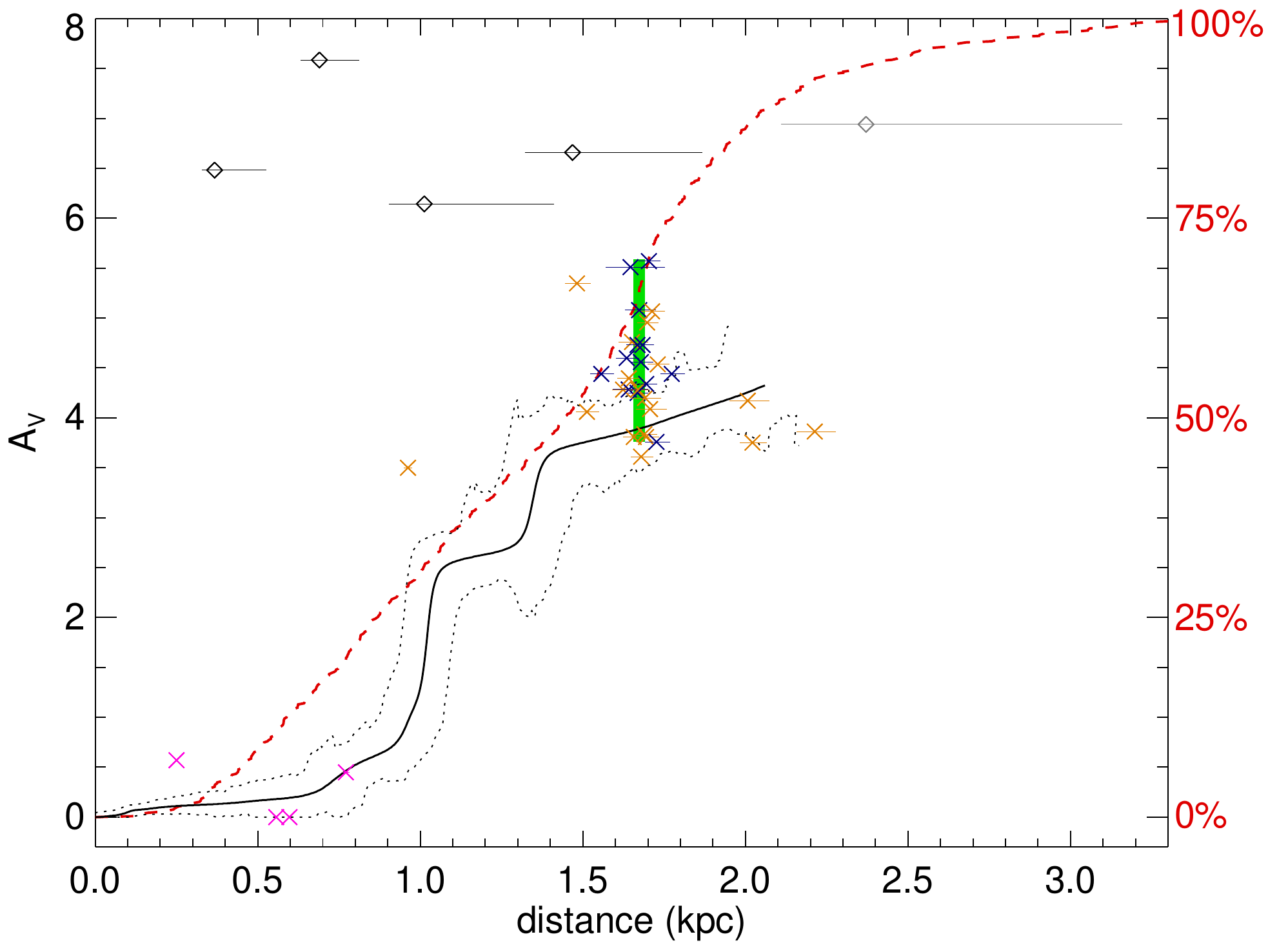}
	\caption{Visual extinction vs. distance estimates, colored as follows: blue, spectroscopically confirmed Berkeley 87 members; orange, photometric candidates for Berkeley 87 membership according to \citet{turner-forbes82}; magenta, additional stars of known spectral types; black, sources not previously cataloged; gray, see below. Objects whose $A_\mathrm{V}$ values are obtained in Sections \ref{sec:individual} and \ref{sec:rjce} are shown as crosses and diamonds, respectively. The Berkeley 87 distance and $A_\mathrm{V}$ range are represented as a green rectangle. The extinction along the Berkeley 87 direction according to the \texttt{Stilism} map is drawn as a black solid curve, with errors shown as dotted lines. The red dashed curve is the cumulative distribution (see scale on the right axis) for the sample of 737 objects with valid distance results, i.e. those that fulfill $0 < f_\varpi \leq 0.2$. The gray data point is not part of this sample and is only included to illustrate the effects of relaxing the $f_\varpi$ constraints (see text). \label{fig:avdist}}
\end{figure}

A total of 737 objects fulfill the $f_\varpi$ requirement in the $15' \times 15'$ OMEGA2000 field, 40 of which have a valid extinction estimate available. The latter are represented in Fig. \ref{fig:avdist}, together with a cumulative histogram of all 737 distances, as well as the $A_\mathrm{V}$ variation along the Berkeley 87 direction according to the \texttt{Stilism} 3-D extinction map \citep{lallement+18}. Our results are consistent with those from \texttt{Stilism}, provided that we take into account that this map may underestimate extinction in directions crossing dense cloud cores \citep[as already warned by][]{capitanio+17}. The abrupt slopes at $\sim 1.0$ and $1.35~\mathrm{kpc}$ may be caused by this condition. The outliers near the upper left corner of Fig. \ref{fig:avdist} will be discussed later.

Fig. \ref{fig:avdist} can be used to test the elimination of KM dwarfs from the results of the RJCE method, as explained in Sect. \ref{sec:rjce}. If the conditions that exclude these nearby red objects ($R_\mathrm{int} > 1.25$ and Equation \ref{eq:rjcelimit}) had been ignored, the region of Fig. \ref{fig:avdist} enclosed by $A_\mathrm{V} > 5$ and $r_\mathrm{est} < 0.8~\mathrm{kpc}$ would have been populated with 21 data points of clearly overestimated extinction. After applying the conditions, only 2 objects are remaining, and their results are close to the imposed limits ($R_\mathrm{int}=1.13, 1.17$, and $A_{K_s}^\mathrm{(RJCE)}=0.74 \pm 0.05, 0.86 \pm 0.08$ for the objects located 367 and 690 pc away, respectively)\footnote{We note that Fig. \ref{fig:avdist} shows two additional RJCE results, located at 1.47 and 1.01 kpc, that might also be overestimated, based on similar arguments (e.g. $R_\mathrm{int}=1.15, 1.24$, respectively). However, evaluation of these cases is less clear, especially because of the more uncertain distances.}. This test shows that our decontamination of RJCE values is effective although not infallible. 

The cumulative distribution shown in red in Fig. \ref{fig:avdist} illustrates that a great majority of measured distances are comparable to the Berkeley 87 distance or shorter. Furthermore, only one out of 737 distances corresponds to a HRP source, implying that \textit{Gaia} EDR3 cannot be used for measuring populations as extinguished as [DB2001] Cl05 (see Sect. \ref{sec:prelim}), at least with the required accuracy. Even if we had relaxed our fractional uncertainty condition up to $f_\varpi = 0.3$, we would have only found 2 HRP objects (one of them whose extinction could be estimated is shown as a gray symbol in Fig. \ref{fig:avdist}), having no evidence of membership to any of the reddened components\footnote{These are two bright objects of nearly identical infrared magnitudes ($K \approx 9.3$) showing marginally HRP colors (both have $J-H \approx 1.5$, $H-K \approx 0.65$), located angularly far ($>5'$) from [DB2001] Cl05, and having no evidence of young age (not detected by \textit{Chandra}, $R_\mathrm{int} \approx 0.7$ for both).} despite their far distances. This shows the importance of complementing \textit{Gaia} distances with extinction results to collect membership evidence throughout the line of sight.
 
\subsubsection{On the Berkeley 87 distance}

Inspection of Fig. \ref{fig:avdist} also reveals that part of the photometric Berkeley 87 candidates listed by \citet{turner-forbes82} have \textit{Gaia} EDR3 distances that are inconsistent with spectroscopic members. The obvious cases whose membership can be fully discarded are the foreground object Be87-54 ($r_\mathrm{est} = 951^{+14}_{-13}~\mathrm{pc}$) and the background stars Be87-5, Be87-7, and Be87-95 (at $2.01^{+0.07}_{-0.06}$, $2.02 \pm 0.04$, and $2.21^{+0.07}_{-0.06}~\mathrm{kpc}$, respectively). In fact, these three background objects might be part of another cluster or association, since they share not only similar parallaxes, but also compatible proper motions.

Since at least part of the \citet{turner-forbes82} candidates are not Berkeley 87 members, we only use spectroscopically confirmed members for measuring the distance to this cluster. We adopt the previously calculated median of 13 hot stars, $r_\mathrm{Be87} = (1673 \pm 17)~\mathrm{pc}$ can be considered as our best distance estimate for Berkeley 87, since no distance results were obtained for any additional spectroscopic members. We note that only the statistical error of the median is considered here.

This result is in excellent agreement with the recent \textit{Gaia} DR2 measurement of 1661 pc obtained by \citet{cantatgaudin+18}, although this result was computed from a different stellar sample and without taking spectral types into account. This agreement slightly contrasts with the 1.75 kpc distance obtained by \citet{skinner+19} as an average of only seven OB stars that were detected in X-rays by \citet{sokal+10}. Anyway, all these \textit{Gaia}-based results are located farther than those from previous spectrophotometric studies in visible wavelengths \citep[who obtained 0.95 and 1.23 kpc, respectively]{turner-forbes82,turner+06}. This discrepancy is probably caused by a slightly overestimated extinction by these authors, who assumed $R_V=3.0$, while we obtain $R_V=2.8$ in Sect. \ref{sec:individual}.

 \section{Classification of young sources}

\subsection{Infrared selection of YSO candidates} \label{sec:iryso}

To find and classify young sources through infrared photometry, the multiband YSO selection criteria by \citet{gutermuth+08, gutermuth+09} (hereafter ``the G09 method'') is commonly applied. This method was designed for a sample of nearby regions undergoing little foreground extinction (in the most obscured cases, comparable to Berkeley 87), and generally located at relatively higher Galactic latitudes. However, the G09 method becomes problematic for observational conditions like those in the ON2 field, which involve selection biases that jeopardize the eficacy of the method. An illustrative example is provided by the spatial distribution [5.8]-band counterparts of highly-reddened \textit{JHK} sources in the Cl05 region (pink crosses in Fig. \ref{fig:cyg05rgb}). While some of these sources can be detected relatively uncrowded regions displaying little extended emission, source confusion makes detection impossible in the densest, most embedded regions. This entails a serious hindrance for phase 1 of the G09 method, which requires simultaneous detection in all IRAC bands. To a lesser degree, this issue is still important in the [3.6] and [4.5] bands, which are simultaneously required for phase 2 of the G09 method.

A detailed characterization of selection biases affecting YSO selection in the infrared (including others that involve Galactic extinction and extragalactic contaminants), is provided in Appendix \ref{sec:bias}. To overcome these biases and produce an acceptably homogeneous census of YSO candidates, we proceed as follows. First, we apply phase 1 of the G09 method, albeit slightly adapted to the observational conditions of ON2. Second, our own method based on the $R_\mathrm{int}$ definition (Sect. \ref{sec:prelim} and Appendix \ref{sec:rint}) is calibrated through results from the first step, and applied to the data instead of phase 2. Finally, supplementary YSO candidates are identified thanks to the far-infrared data.

\subsubsection{IRAC-based selection} \label{sec:phase1}

First of all, we have tested phase 1, as originally published by \citet{gutermuth+09}, on the $15' \times 15'$ OMEGA2000 field, yielding 63 YSO candidates and 31 contaminants. Among the latter, 19 were categorized as extragalactic objects, of which 14 were classified as PAH galaxies, and 5 as Broad-Line Active Galactic Nucleus (BL-AGN). However, the ratio of extragalactic contaminants to YSO candidates seems unrealistically high for a region experiencing clustered star formation in the Galactic plane, where extragalactic light is extremely extinguished after crossing a long path throughout the Milky Way. Indeed, a significant amount of YSOs are being misidentified as PAH or BL-AGN galaxies, owing to magnitude cuts in the G09 method ($[4.5]_\mathrm{PAH}<11.5$, and $[4.5]_\mathrm{BL{\text{-}}AGN} < 13.5$, respectively) that correspond to the YSO population in the nearest kiloparsec.

In order to correct these misidentifications, we modify the aforementioned magnitude cuts as fully explained and justified in Appendix \ref{sec:extbias}. Briefly, the adaption needed consists of changing the BL-AGN limit one magnitude fainter ($[4.5]>14.5$), and finding that only $\sim 0.64$ PAH galaxies are expected in the OMEGA2000 field in the range $12.5<[4.5]<15$, which is shared by YSOs. We consider this figure low enough to simply ignore extragalactic contamination behind ON 2S. 

We adhere to the remaining steps of phase 1 (namely: removal of shock emission knots and PAH-contaminated apertures; selection of class 0/I candidates; and the same for class II), as in the original G09 method.

Among the results, we notice that WR 142 and three Be-type members of Berkeley 87 (specifically V439 Cyg, VES 203, and VES 204) are categorized as class II candidates. This is not totally unexpected, since Wolf-Rayet and Be stars commonly show strong mid-infrared excesses \citep{mauerhan+11}, originated in their dense, extended envelopes \citep{gehrz+74,cohen+75}, whose IRAC colors can resemble those from YSOs. We manually remove these four stars from the YSO candidate list.

Our adapted version of phase 1 finally yield 46 class 0/I candidates, 26 class II candidates, and 15 PAH-contaminated apertures. In the latter case, accidental PAH contamination cannot be distinguished from that of circumstellar origin. This ambiguity was not important for the observational sample of \citet{gutermuth+09}, dominated by T Tauri stars, which rarely show PAH emission \citep{furlan+06,hernandez+07}. However, it becomes relevant for Herbig Ae/Be stars, which commonly display strong PAH emission \citep{acke-vandenancker04,chen+16}. In any case, we prefer not to jeopardize the reliability of our YSO list (which is crucial for the next section), and therefore we keep considering these 15 objects as contaminants. 

\subsubsection{$R_\mathrm{int}$-based selection} \label{sec:rintyso}

We aim at extending YSO classification to sources that are not detectable in the longest IRAC wavelengths, in a similar way as phase 2 of the G09 method, but managing the involved selection effects (Appendix \ref{sec:photbias}). The main idea behind phase 2 of the G09 method is distinguishing intrinsic reddening from color excess caused by interstellar extinction, based on the \textit{JHK}[3.6][4.5] colors of the reddened sources relative to the corresponding extinction vectors. The intrinsic reddening index ($R_\mathrm{int}$) defined in Sect. \ref{sec:intred} and Appendix \ref{sec:rint}, is designed for the same purpose, although offering the following advantages. First, $R_\mathrm{int}$ can be still calculated when any one of the bands (except $H$) is missing, whereas only $J$ can be waived for the classical phase 2; this is especially relevant at the southwestern part of the OMEGA2000 that lack $4.5~\mu \mathrm{m}$ observations (Fig. \ref{fig:coverage}). Second, nearly all $R_\mathrm{int}$ values are computed by averaging results from 2 or more (and usually all 6) of the color-color combinations used in its definition (Appendix \ref{sec:rint}); this allows to relax the uncertainty constraints required for the G09 method ($\sigma_\mathrm{JHK}<0.1$; $\sigma_\mathrm{IRAC}<0.2$). This flexibility results in a significant increase of point sources that can be evaluated; quantitative details are discussed in Appendix \ref{sec:assessable}.

\begin{figure}
	\centering
	\includegraphics[width=0.38\textwidth, bb=35 7 295 211]{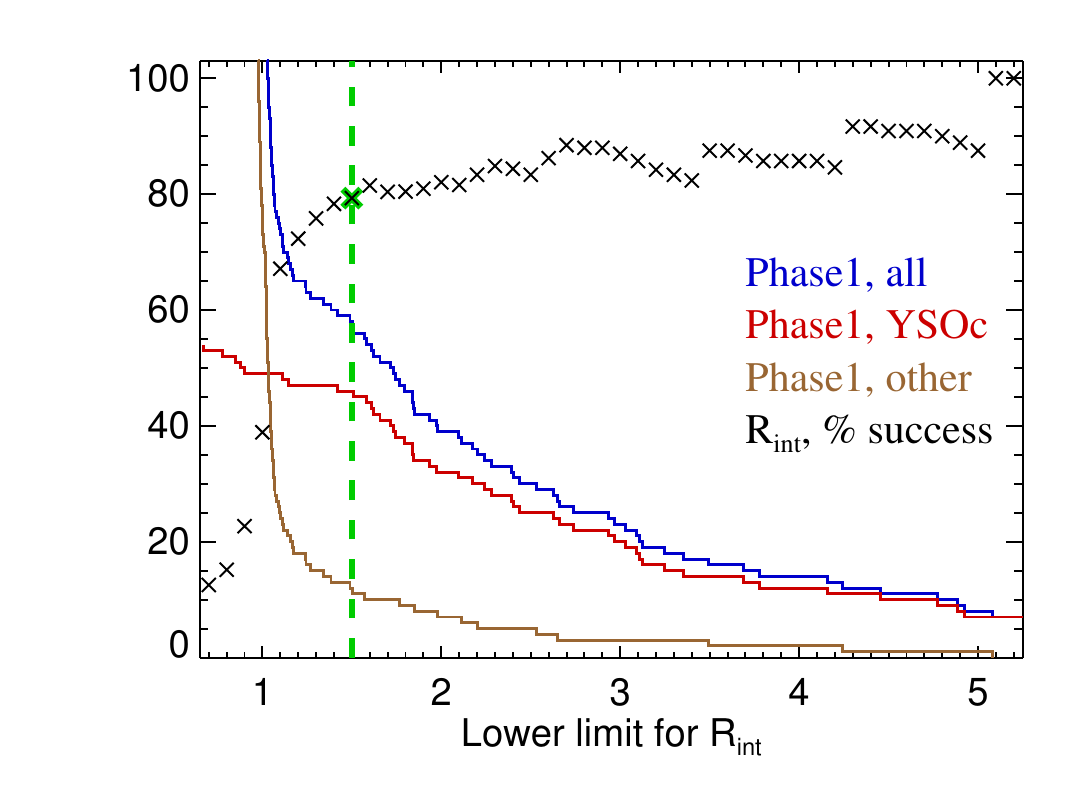}
	\caption{Calibration of the $R_\mathrm{int}$ method through the sample and results of adapted phase 1; see text for explanation. The vertical axis represents either the number of objects for the colored lines, or percentages for the black crosses. The reader should be careful not to interpret horizontal axis as $R_\mathrm{int}$ values, but as $R_\mathrm{int}$ ranges. The finally selected criterion, $R_\mathrm{int} > 1.5$, is marked in green. \label{fig:rintyso}}
\end{figure}

These arguments lead us to propose a new method for identifying additional YSO candidates based on the intrinsic reddening index (hereafter, ``The $R_\mathrm{int}$ method''). It consists of finding the value $m$ that makes the condition $R_\mathrm{int}>m$ optimal for YSO selection. The success rate (i.e. the proportion of genuine YSOs among the selected sources) is expected to grow with $m$, although at the expense of leaving out an increasing number of candidates with mild infrared excess. Therefore, we have to find a compromise between the amount of resulting YSO candidates and an acceptably high success rate. For this purpose, we take the sample analyzed by the adapted phase 1 (i.e. all point sources detected in all IRAC bands), and the resulting YSO candidates (Sect. \ref{sec:phase1}) are assumed to be true. The behaviour of phase 1 YSOs that are recovered by the $R_\mathrm{int}$ is shown in Fig. \ref{fig:rintyso}. For low $m$ values, the success rate rises steadily as few YSOs escape from the $R_\mathrm{int}>m$. This trend changes abruptly at $m \approx 1.5$, where the amount of recovered YSOs begin to fall more steeply than false positives. From that point, success rate variations are not significantly, and eventually become stochastic due to low number statistics.

On this basis, we choose $R_\mathrm{int}>1.5$ as the optimal criterion for selecting additional YSO candidates. The green cross in Fig. \ref{fig:rintyso} indicates that $\sim 80\%$ of objects fulfilling the condition $R_\mathrm{int}>1.5$ are expected to be real YSOs, provided that extragalactic contamination is negligible (which is justified in Appendix \ref{sec:extbias}). A plausible source of contamination is caused by nearby ($<0.6~\mathrm{kpc}$) low-mass dwarfs whose $R_{int}$ would be slightly overestimated. However, only 4 objects fulfill $R_\mathrm{int}>1.5$ and $r_{est}<0.6~\mathrm{kpc}$ simultaneously, and all but one are too bright to be main-sequence stars. Therefore, red dwarfs represent a minor contribution to $R_\mathrm{int}$ overestimation.

After excluding sources that have been classified in previous sections (either as YSO candidates or as contaminants), the $R_\mathrm{int}$ method yields 290 new YSO candidates, of which $~58$ (20\%) are expected to be false positives. Among the 290 new candidates, 94 (8 of which are projected on the Cl05 region) would have never been evaluated with the classical G09 method (including its phase 2; see details in Appendix \ref{sec:assessable}).

To ensure that fake YSO candidates are not produced by high uncertainties, we double-check that inaccurate detections are properly compensated by averaging 2 or more color-color combinations. Among the 290 new candidates, only 18 $R_\mathrm{int}$ values are produced through a single color-color combination, and all of them are photometrically accurate enough to fulfill the uncertainty requirements for the original phase 2, except one having $R_\mathrm{int}=6.06$, which is high enough to claim that its large infrared excess is authentic.

\subsubsection{Far-infrared point sources}

To search for additional objects in early phases of star formation, especially those that are too embedded to be detected in the mid-infrared, we examine the list of \textit{Herschel}/PACS point-source detections.

Within the OMEGA2000 field, 11 point sources are detected in the PACS $70~\mu \mathrm{m}$ band, which is considered as an excelent indicator of the presence of a protostar \citep{dunham+08,konyves+15}. These include the source that corresponds to the Cygnus 2N massive star-forming region (already addressed in Sect. \ref{sec:gaia}), as wfell as 5 YSO candidates found through the adapted phase 1 (Sect. \ref{sec:phase1}), and 1 LRP object that was discarded as a PAH-contaminated aperture (also in Sect. \ref{sec:phase1}). Another $70~\mu \mathrm{m}$ detection, whose IRAC counterpart is the brightest in the field (e.g. $[5.8]=3.93 \pm 0.02$), cannot be considered as an YSO, since it is a carbon star according to \citet{alksnis+01}. We classify the 3 remaining sources as new candidate protostars. One of them is detected in two IRAC bands, being the corresponding color ($1.03 \pm 0.012$) consistent with a class 0/I object.

On the other hand, 27 sources are detected only in the $160~\mu \mathrm{m}$ band. We consider these objects as candidate starless cores based on \cite{ragan+12,lippok+16} \citep[see however][]{feng+16}.

\begin{table*}
\caption{Catalog of sources with extinction, distance, or object classification results} \label{tab:results}
\centering
\begin{tabular}{rcclrlllrrcl} 
\hline\hline
 Src &	R.A. 		 &	 Dec  	 & Literature 		       &  $K$  & Matched 		  & Pop. & Cl05 & $R_\mathrm{int}$ & $A_v$\tablefootmark{d} & $r_\mathrm{est}$ & Source     \\
 \#  & ($20^\mathrm{h}$) & ($+37^\circ$) & identifier\tablefootmark{\;a} & (mag) & observ.\tablefootmark{b} &      & reg.\tablefootmark{c} & 	 	    & (mag) & (kpc) & classif.\tablefootmark{e} \\
\hline
 1  &  21:15.34 & 24:30.9 & HD 229059  & 4.865 &   CGOS & LRP  &   out & 0.991 &      5.08   &$1.67_{-0.04}^{+0.05}$    & B2Iabe      \\
 2  &  21:23.80 & 20:01.4 &            & 6.290 &   GOS  & LRP  &   out & 0.848 &             &$1.10_{-0.02}^{+0.02}$    &             \\
 3  &  21:56.20 & 21:28.5 & HD 229105  & 6.886 &   GOS  & LRP  &   out & 0.780 &      0      &$0.597_{-0.004}^{+0.004}$ & K2II        \\
 4  &  22:14.61 & 16:14.7 &            & 6.940 &   GOS  & LRP  &   out & 0.757 &             &$3.15_{0.22}^{0.39}$      &             \\
 5  &  21:35.54 & 23:29.6 & BD+36.4031 & 7.107 &   GOS  & LRP  &   out & 0.787 &      0      &$0.557_{-0.004}^{+0.004}$ & K0III       \\
 6  &  21:34.04 & 23:53.1 &            & 7.264 &   GOS  & LRP  &   out & 0.731 &             &$1.80_{-0.07}^{+0.10}$    &             \\
 7  &  21:46.38 & 15:37.0 &            & 7.334 &   GOS  & LRP  &   out & 1.021 &             &$1.44_{-0.03}^{+0.04}$    &             \\
 8  &  21:50.52 & 28:57.0 &            & 7.411 &   GOS  & LRP  &   out & 0.794 &             &$1.09_{-0.02}^{+0.02}$    &             \\
 9  &  21:38.66 & 25:15.0 & BD+36.4032 & 7.455 &   CGOS & LRP  &   in  & 1.000 &      4.34   &$1.69_{-0.03}^{+0.03}$    & O8.5-9III-V \\
 10 &  21:33.56 & 24:51.4 & V439 Cyg   & 7.629 &   GOS  & LRP  &   out & 1.489 &      4.28   &$1.64_{-0.05}^{+0.06}$    & B1.5Ve      \\
 11 &  21:37.02 & 24:17.1 &            & 7.811 &   OS   & HRP  &   out & 0.735 & {\it 19.45} &                          &             \\
 12 &  22:07.64 & 28:12.3 &            & 7.845 &   GOS  & HRP  &   out & 0.717 & {\it 7.12}  &                          &             \\
  ...   &  ... & ...  &  ...    & ... &   ...     & ...  &   ...   &   ...  &      ...    &   ...  & ...     \\
\hline
\end{tabular}
\tablefoot{The full version of this table is available online at the CDS. Objects are sorted by K-band magnitude, or by R.A., when $K$ is missing.\\
\tablefoottext{a}{This identifier can be either: a SIMBAD ID; or ``Cygnus 2N'' and ``Cygnus 2N EAST'' for the so-called components as described in Sect. \ref{sec:cyg2n}; or ``$\mathrm{Sk19~T}a{\text{-}}bb$'', standing for \citet[Table $a$, line $bb$]{skinner+19}.}
\tablefoottext{b}{List of observations where a counterpart has been found, coded as: C=\textit{Chandra}, likely counterpart; c=\textit{Chandra}, doubtful counterpart; G=\textit{Gaia}; O=OMEGA2000 (even when 2MASS is used for photometry instead); S=\textit{Spitzer}; H=\textit{Herschel}.}
\tablefoottext{c}{Source inside or outside the Cl05 region (as defined in Sect. \ref{sec:prelim}).}
\tablefoottext{d}{Values computed through the RJCE method are shown in italics.}
\tablefoottext{e}{This column shows either the known spectral type, or our classification as YSO candidate according to the following abbreviations: \textit{cI}, class I; \textit{cII}, class II; \textit{cIII}, class III; \textit{c.protost.}, candidate protostar; \textit{c.starless}, candidate starless core; \textit{massiveSF}, massive star-forming region; \textit{othX}, other X-ray-detected candidate; \textit{Rint}: identified through the $R_\mathrm{int}$ method; \textit{YSOc}, YSO candidate.}
}

\end{table*}

\subsection{X-ray emitting sources} \label{sec:xclass}

Along pre-main sequence evolution, X-ray luminosity decays much more slowly than infrared excess \citep{preibisch-feigelson05}, making YSOs still detectable in X-rays even after they become indistinguishable from main-sequence stars in the infrared, i.e. when they become class III sources. For this reason, X-rays are crucial for finding class III YSOs, as shown by many recent observational studies \citep[e.g.][]{wang+08,wang+09,stelzer+12,romanzuniga+15,riveragalvez+15}. On the other hand, X-ray emission is often observed in hot luminous stars \citep{feigelson+07} with lifetimes of few Myr \citep{georgy+12}. Therefore, X-ray point sources are ideal to pinpoint members of any young population regardless of their evolutionary state, from YSOs to Wolf-Rayet stars.

First of all, we exclude X-ray data without reliable infrared counterparts, since most of them are expected to be produced by shocks or background AGN galaxies \citep[see][]{getman+05}. Among the 200 likely counterparts of \textit{Chandra} sources (Sect. \ref{sec:gaia}), 7 are spectroscopic members of Berkeley 87, 1 is the foreground late-type star TYC 2684-25-1 (F8V, $r_\mathrm{est}=(250 \pm 1)~\mathrm{pc}$); and 17 have been classified as YSO candidates in Sect. \ref{sec:iryso} (specifically, 10 through the adapted phase 1, and 7 through the $R_\mathrm{int}$ method). The 175 remaining X-ray emitters with reliable infrared counterparts are tentatively classified as class III when $R_\mathrm{int} \leq 1.5$, or as other X-ray-detected YSO candidates when $R_\mathrm{int}$ could not be calculated.

Still, a non-negligible amount of field stars are expected to contaminate our X-ray sample. Fortunately, the strong decay in X-ray luminosity for stars older than $\gtrsim 10^8~\mathrm{yr}$ \citep{micela+93,randich00,preibisch-feigelson05}, as well as for supergiants of spectral type later than B1 \citep{berghofer+97,clark+19}, limits this field-star contamination to foreground stars, like the aforementioned case of TYC 2684-25-1. This claim is supported by the \textit{Gaia} distance distribution for \textit{Chandra} sources (a total of 41, taking only $f_\varpi \leq 0.2$ sources; see Sect. \ref{sec:Be87}). This distribution is clearly bimodal: 9 objects are closer than 0.65 kpc, and all but two of the remaining sources are farther than $1.35~\mathrm{kpc}$ are farther (unsurprisingly peaking at the Berkeley 87 distance). We note  that such bimodality is not observed in the cumulative histogram of Fig. \ref{fig:avdist}. Consequently, we have decided to remove all 10 X-ray sources that closer than 1 kpc from the YSO candidate list. As a result, 108 class III objects, along with 58 other YSO candidates identified through X-ray emission are added to our list of candidates for young population membership.

\section{The catalog: discussion and future work}

Among the 47090 entries of our multi-wavelength point-source catalog, extinction has been estimated for 1823 objects ($\approx 3.9\%$), distance has been calculated for 737 stars ($\approx 1.6\%$) with accurate \textit{Gaia} EDR3 parallaxes, and 571 sources ($\approx 1.2\%$) have been classified as object types that are compatible with recent or ongoing star formation (hot stars, YSO candidates, dense cores). As a whole, the 3005 objects that are presented in Table \ref{tab:results} have at least one of these results determined. Further astrometric and photometric data for these sources are presented in Appendix \ref{sec:fulldata}.

We consider that at least one of the aforementioned properties (extinction, distance, classification) is required to potentially determine membership to one of the young populations that are overlapped along the line of sight. We note that proper motions are not listed as they have to be complemented with parallax information in this Galactic direction where dependence with distance is weak (see Sect. \ref{sec:intro}). Such task is beyond the scope of the present work and will be presented in the next paper of this series.

Having distance information (obtained directly from parallaxes or indirectly from extinction) would be desirable for every object whose classification make it a suspected member of a young population. However, only 86 out of the 571 sources thus classified have extinction or distance estimates available. Even worse, among YSO candidates located within the Cl05 region (71 in total), distance is only determined in 3 cases, being compatible with Berkeley 87 or nearbier, while extiction is estimated for another 1. This lack of extinction and distance results for the most relevant objects is caused by the following selection efects. First, YSOs are not easily detected in optical wavelengths, and specifically by \textit{Gaia}, due to intrinsic reddening and extinction (especially in the case of ON2). Second, the majority of YSO candidates are intrinsically redder than $R_\mathrm{int} = 1.25$, which bans them from the RJCE method (see Sect. \ref{sec:rjce}). Third, detection in the IRAC $4.5~\mu m$ band, which is required for the RJCE method, is significantly affected by the photometric incompleteness effects described in Sect. \ref{sec:intred} and Appendix \ref{sec:photbias}.

Owing to these difficulties, further analysis is required to effectively establish membership and separate the overlapped young populations, which may include not only the two clusters, but also young field stars. These objectives will be addressed in the next paper of this series, where the practical definitions established in Sect. \ref{sec:prelim} and the catalog published in Tables \ref{tab:results} and \ref{tab:fulldata} will be taken as a basis. This future work will also include other analyses that have been explicitly postponed herein, e.g. on X-ray fluxes or proper motions.

\section{Conclusions}

We have developed a new methodology for building a census of young objects in cases where distinct young populations are overlapped along the line of sight. In such cases, point sources are observed under a variety of distances and extinction conditions, and our flexible approach has the ability to overcome the involved selection effects.

The cornerstone of our methodology is the intrinsic reddening index, $R_\mathrm{int}$, which separates intrinsically red objects from those whose color excess is caused by interstellar extinction. Unlike other previously known techniques, $R_\mathrm{int}$ is designed to work under conditions of significant photometric incompleteness. We have shown that, in such situation, the usefulness of $R_\mathrm{int}$ is twofold. On the one hand, it allows to avoid intrinsically red objects when using extinction estimation methods that are solely based on photometry (e.g. the RJCE method), since these objects would yield overestimated values. On the other hand, $R_\mathrm{int}$ can be used to pinpoint YSOs that would be otherwise neglected due to incomplete photometry. Based on the latter, a new method for YSO candidate selection is presented in Sect. \ref{sec:rintyso}.

Our methodology is applied to multi-wavelength observations (from the far infrared to X-rays) of a field where cluster formation has taken place at different distances. As a result, 571 point sources are classified as objects related to recent or ongoing star formation, with evolutionary stages ranging from starless cores to evolved hot massive stars. These include 290 YSO candidates that have been found thanks to the $R_\mathrm{int}$ method, of which $\sim 80\%$ of them are expected to be real YSOs.

Tables \ref{tab:results} lists not only objects whose classification is compatible with a young population, but also other unclassified sources whose extinction or distance estimates can lead to membership determinations. Hence, the methods and results presented here will allow to disentangle the overlapped populations and further characterize the Berkeley 87 and [DB2001] Cl05 clusters in a forthcoming paper.

\begin{acknowledgements}

We thank the anonymous referee for her/his constructive comments that improved the manuscript. We are grateful to Luis Aguilar for useful suggestions about \textit{Gaia} parallaxes. D.dF. acknowledges the UNAM-DGAPA postdoctoral grant, the Generalitat Valenciana APOSTD/220/228 fellowship, and the Spanish Governent AYA2015-68012-C2-2-P project. D.dF. and M.G. have received financial support from Spanish Government project ESP2017-86582-C4-1-R. C.R.-Z. acknowledges support from CONACYT project CB2017-2018 A1-S-9754. C.R.-Z. and E.J.B. acknowledge support from Programa de Apoyo a Proyectos de Investigaci\'on e Innovaci\'on Tecnol\'ogica, UNAM-DGAPA, grants IN108117 and IN109217, respectively. The scientific results reported in this article are based in part on data obtained from the Chandra Data Archive. This research has made use of software provided by the Chandra X-ray Center (CXC) in the application package CIAO. This publication makes use of data products from the Two Micron All Sky Survey, which is a joint project of the University of Massachusetts and the Infrared Processing and Analysis Center/California Institute of Technology, funded by the National Aeronautics and Space Administration and the National Science Foundation. Herschel is an ESA space observatory with science instruments provided by European-led Principal Investigator consortia and with important participation from NASA. This work has made use of data from the European Space Agency (ESA) mission {\it \textit{Gaia}} (\url{https://www.cosmos.esa.int/gaia}), processed by the {\it \textit{Gaia}} Data Processing and Analysis Consortium (DPAC, \url{https://www.cosmos.esa.int/web/gaia/dpac/consortium}). Funding for the DPAC has been provided by national institutions, in particular the institutions participating in the {\it \textit{Gaia}} Multilateral Agreement. This research has made use of the SIMBAD database, operated at CDS, Strasbourg, France.

\end{acknowledgements}

%
   \bibliographystyle{aa} 
   \bibliography{cyg05paper1} 
%

\begin{appendix}

\section{Details on the intrinsic reddening index} \label{sec:rint}

As introduced in Sect. \ref{sec:intred}, $R_\mathrm{int}$ is defined in terms of the position of the sources in color-color diagrams whose horizontal and vertical axes are, respectively, a mid-infrared color $X=X_2-X_1$, and a near-infrared color $Y=Y_2-Y_1$. We avoid colors that include the longest IRAC wavelegths ([5.8] and [8.0]) because the small number of detections in these bands would make them impractical for a relatively homogeneous study of the whole stellar population. Regarding the near-infrared colors, $J-H$ is, in principle, the most suitable to isolate interstellar reddening, as in \citet{teixeira+12}. However, the fact that $J$ is missing in a large fraction ($\approx 14\%$) of sources detected in $H$ and $K$ has lead us to include $H-K$ as well. Therefore, the following options for $X$ and $Y$ are allowed: $X=K-[3.6]$, $X=K-[4.5]$, $X=[3.6]-[4.5]$; and $Y=J-H$, $Y=H-K$.

\begin{figure*}
	\centering
	\includegraphics[width=\textwidth]{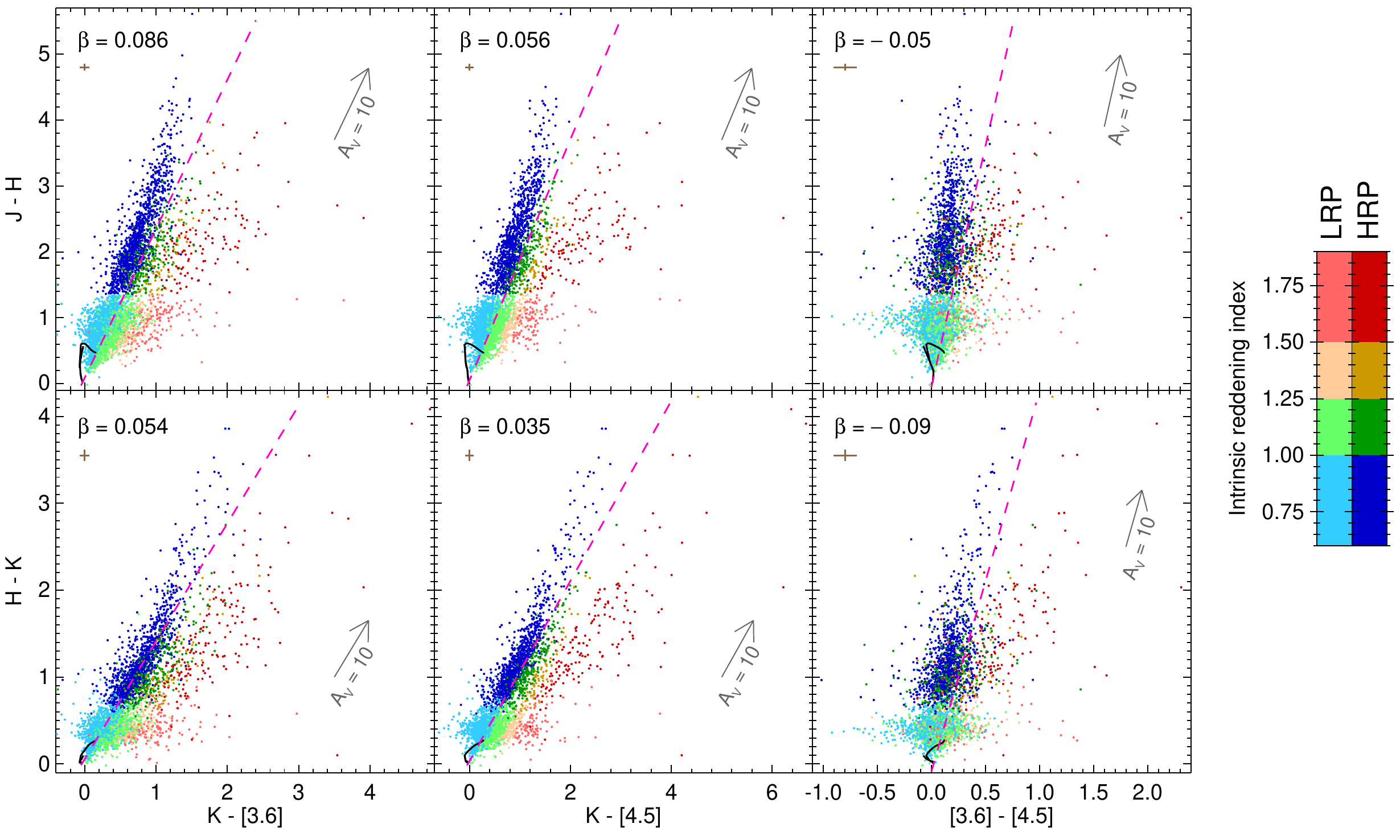}
	\caption{Color-color diagrams that are used to compute $R_\mathrm{int}$, which is coded as indicated in the color bars for LRP and HRP sources. The reddening band boundaries that are used as a reference (see text) are drawn as pink dashed lines. The 1 Gyr isochrone for solar abundances is drawn in black. The average photometric errors are shown as brown bars just below the $\beta$ values. \label{fig:rintccd}}
\end{figure*}

In the six resulting color-color diagrams (Fig. \ref{fig:rintccd}), the extinction vector is drawn according to the \citet{indebetouw+05} law. Since the $A_K$ value provided by these authors actually refers to the 2MASS $K_S$ band, we apply the correction $A_K/A_{K_S}=0.98$, that we obtain through approximate interpolation of the \citet{indebetouw+05} law at the effective wavelength of the UKIDSS $K$ bandpass.

The next step is to determine, in each diagram, the boundary between the interstellar reddening band of normal stars and the region where intrinsically red stars should be placed. More specifically, the boundary where, for each given $Y$ value, $X$ becomes too red to be consistent with a normal star that is only affected by interstellar extinction. To represent a typical population of unreddened stars (like those found throughout the Galactic disk outside star-forming regions), a 1 Gyr isochrone for solar abundances is taken from the Dartmouth Stellar Evolution Database \citep{dotter+08}. In each diagram, normal stars are expected to lie within the reddening band that results from sliding the isochrone (shown as a black curve) along the direction of the extinction vector. The limit of the reddening band with the reddest $X$ values (although not strictly, as will be clarified below) is the boundary we are looking for. In each Fig. \ref{fig:rintccd} diagram, the boundary (pink dashed line) follows the equation:

\begin{equation} Y = \alpha(X,Y) \cdot X + \beta(X,Y) \;, \quad \mathrm{where}~\alpha(X,Y)=\frac{A_{Y_2}-A_{Y_1}}{A_{X_2}-A_{X_1}}\,,
 \label{eq:reddening}
\end{equation}
and $\beta(X,Y)$ takes the value that is printed in the corresponding pannel of Fig. \ref{fig:rintccd}. A strict compliance with this boundary definition would imply that all the isochrone data points would have bluer $X$ colors than the boundary. In some cases, however, the sole contribution of low-mass objects, which are not expected to be detectable at Cygnus-X distances, pushes the boundary towards significantly redder colors. To avoid this undesirable effect, the low-mass end of the isochrone is ignored in such cases, and any possible contamination in the foreground (where low mass stars are detectable) is assessed in Sect. \ref{sec:rintyso}.

Then, for each diagram $(X,Y)$, the relative shift from the reddening band boundary is defined as:

\begin{equation}
 S_{X,Y} = \frac{X-\left[Y-\beta(X,Y)\right]/\alpha(X,Y)}{\log (\lambda_{X_2} / \lambda_{X_1})} \,,
 \label{eq:hshift}
\end{equation}
and $\lambda$ stands for the effective wavelength of the corresponding filter. We remark that the shift is defined as relative to the corresponding color baseline, $\Delta \log \lambda$.

Finally, $R_\mathrm{int}$ is computed as a weighted average of relative shifts as expressed in equation \ref{eq:rint}. This $R_\mathrm{int}$ definition implies that a hypothetical data point that is simultaneously located on all the reddening band boundaries would yield $R_\mathrm{int}=1$. The weights in equation \ref{eq:rint} can be tailored to the scientific goals that are addressed; in this work we have chosen the following weights that depend on the relative photometric uncertainties: 

\begin{equation}
 W_{X,Y} = \frac{\log(\lambda_{X_2}/\lambda_{X_1})}{\sqrt{(\sigma_{X_1}^2+\sigma_{X_2}^2)+(\sigma_{Y_1}^2+\sigma_{Y_2}^2)/\alpha(X,Y)}}
\end{equation}

\section{Description and treatment of relevant selection biases} \label{sec:bias}

  \subsection{Purely photometric biases affecting the G09 method} \label{sec:photbias}

\begin{figure}
	\centering
	\includegraphics[width=0.38\textwidth]{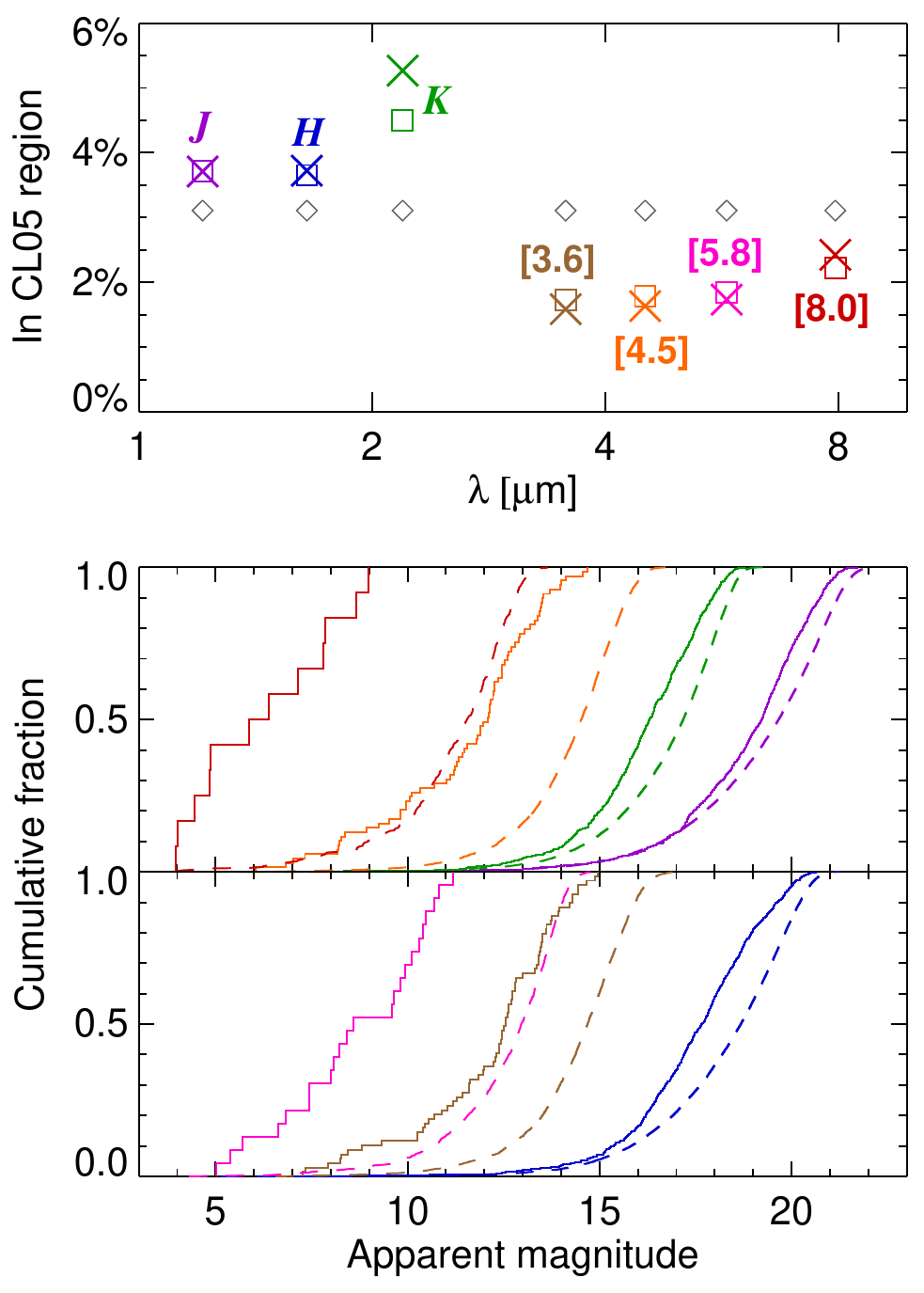}
	\caption{Illustration of selection biases affecting infrared photometry in the Cl05 region, relative to the NIR+MIR region. \textit{(Top)} Proportion of detections falling within the Cl05 region, as computed by considering all detections (squares), or only those meeting the corresponding uncertainty requirement for the G09 method (crosses). For comparison, the case of spatially uniform distribution is shown as gray diamonds. \textit{(Bottom)} Cumulative histograms of apparent magnitudes in the Cl05 region (solid lines) and the whole NIR+MIR region (dashed), using the same uncertainty-limited sample as for cross symbols in the top pannel, as well as the same color code. \label{fig:cl05bias}}
\end{figure}

Phase 1 of the G09 method requires that sources are simultaneously detected in all four IRAC bands with photometric uncertainties lower than 0.2, while only the [5.8] and [8.0] bands can be waived for phase 2 (which in return needs at least $H$ and $K$ with $\sigma<0.1$). We remark that these conditions are required for being merely evaluated by the method, before applying any classification criteria. Our data are affected by these restrictions in the following ways. First they would confine our YSO search to the portion of the OMEGA2000 field that is covered by all the four IRAC bands (hereafter ``the NIR+MIR region''; see also Fig. \ref{fig:coverage}), spanning roughly $158~\mathrm{arcmin}^2$. Second, the density of detections meeting the phase 1 requirements decreases toward the Cl05 region (as defined in Sect. \ref{sec:prelim}), despite the near-infrared overdensity. The latter is better illustrated in Fig. \ref{fig:cl05bias}, which shows the amount and distribution of photometric detections in each band for the Cl05 region, relative to the NIR+MIR region.

The top pannel of Fig. \ref{fig:cl05bias} reveals that the overdensity of near-infrared sources at the Cl05 region, which is especially conspicuous in the $K$ band, turns into an underdensity in IRAC wavelengths, where YSOs were supposed to appear highlighted. This problem is caused by the combined effects of high distance towards ON2 \citep[relative to the regions studed by][]{gutermuth+08,gutermuth+09}, and the IRAC technical features. On the one hand, stellar crowding in [DB2001] Cl05 appears to be excessive for the limited spatial resolution of IRAC (compared to OMEGA2000), which leads to strong source confusion effects. On the other hand, the presence of distant bright clouds aligned with point sources makes the latter hard to detect due to low contrast\footnote{For stars embedded in a cloud, contrast significantly worsens with the observer's distance. While the measured flux of a point source decreases as $1/r^2$, the surface brightness of the cloud remains constant (since the covered solid angle also decreases as $1/r^2$).} in the IRAC images, especially in the bands that are more affected by Polycyclic Aromatic Hydrocarbon (PAH) emission, [5.8] and [8.0]. Still, a slight increase towards the longest wavelengths is noticeable in the plot, which is interpreted as the presence of disks slightly compensating the selection bias.

The above explained biases are not only detrimental to source counts, but also to the magnitude ranges where the G09 method can be used, as shown in the lower plots of Fig. \ref{fig:cl05bias}. Cumulative histograms therein reveal that the G09 method is biased against faint objects in the Cl05 region, in comparison with the NIR+MIR field. Although this occurs in every photometric band, the differences between both samples are larger for the IRAC photometry, and especially in the [5.8] and [8.0] bands, where only detections in the bright tail of the distribution survive contamination from extended PAH emission.

Altogether, the original G09 method can be applied to 3718 sources, of which 58 (1.56\%) are part of the Cl05 region. If we only focus on phase 1, 480 sources are assessable in total, 10 of them (2.1\%) inside the Cl05 circle. Despite the overdensity of red objects, these proportions are notably lower than the corresponding ratio of solid angles, 3.1\%, that would match a spatially uniform distribution. Indeed, a strict application of the G09 method would miss a very significant part of red sources just in the region where YSOs are expected to be concentrated.

\subsection{Extinction-related biases} \label{sec:extbias}

To distinguish YSO candidates from extragalactic contaminants, the original G09 method included constraints in [4.5] (namely: $[4.5]_\mathrm{PAH}>11.5$; $[4.5]_\mathrm{BL{\text{-}}AGN}>13.5$) that were based on the apparent magnitude distribution of YSOs in the nearest kiloparsec \citep{gutermuth+08}. In the ON2 region, however, we expect to find YSOs about 4 times that distance, implying a brighness decrease of $\sim 3$ magnitudes. Another half magnitude should be added to [4.5] because of the additional foreground extinction \citep[$\Delta A_\mathrm{V} \sim 10$, converted through][]{rieke-lebofsky85,indebetouw+05}, relative to the \citet{gutermuth+09} sample. As a consequence, the G09 method would have erroneously discarded any YSOs that fall within a 3.5 magnitude range from the original limits (i.e. $11.5<[4.5]_\mathrm{PAH}<15$; $13.5<[4.5]_\mathrm{BL{\text{-}}AGN}<17$).

This issue cannot be simply solved by shifting the [4.5] boundaries accordingly, since galaxies might be still present as contaminators whithin that 3.5 magnitude range. Extragalactic light is attenuated by interstellar extinction while traveling throughout the Galactic disk, starting from some point far beyond ON2. The total extinction must be higher than the most extinguished stellar populations observed in off-cloud regions. i.e. a few tens of visual magnitudes (Based on the extent of color-color diagrams along the extinction vector, see e.g. Fig. \ref{fig:cyg05rgb}b). Taking a conservative approach, w assume $\Delta A_{[4.5]}=1$ (roughly equivalent to $\Delta A_\mathrm{V} \approx 20$) for extragalactic sources \citep[relative to the negligible extinction of the Bootes field used by][]{gutermuth+08}. In our field, the 5 sources initially categorized as BL-AGN candidates are in the range $13.5<[4.5]<14.5$, therefore this 1-magnitude dimming is enough to simply skip the step of finding BL-AGN contaminants.

The case of extragalactic PAH candidates is more complicated, since only 3 out of the 14 initially selected sources are in the range $11.5<[4.5]<12.5$. To assess the degree of contamination by PAH galaxies still present among the 11 remaining objects, we rely on the power law that \citet{gutermuth+08} fitted to the differential surface density of PAH galaxies ($\sigma_\mathrm{PAH} ([5.8])$) in the Bootes field. If we assume a representative color $[4.5]-[5.8] \approx 0.5$ for PAH galaxies, based on \citet[Fig. 13]{gutermuth+09}, and we take into account the extra extinction for the ON2 direction relative to Bootes, then:

\begin{eqnarray}
 \sigma_\mathrm{PAH}^\mathrm{(ON2)} (12.5<[4.5]<15) \approx \sigma_\mathrm{PAH}^\mathrm{(Boo)}(11<[5.8]<13.5) \nonumber \\
 = 2 \int_{11}^{13.5} 10^{-18.24} [5.8]^{17.02}~\mathrm{d}[5.8] = 14.6~ \mathrm{deg}^{-2}\,,
 \label{eq:pah}
\end{eqnarray}
\noindent
where the factor 2 takes into account that $\sigma_\mathrm{PAH} ([5.8])$ is defined per half-magnitude bin \citep{gutermuth+08}. Equation \ref{eq:pah} implies that the expected number of PAH galaxies within the solid angle covered by the NIR+MIR field is 0.64, or equivalently, that the 11 aforementioned sources have a $\sim 7\%$ probability of being a PAH galaxy each. Therefore, if we allow some likelihood of 1 or 2 PAH galaxies being still misidentified as YSOs, we can skip the elimination of PAH galaxies as we did for BL-AGN candidates above.

Finally, we have to ensure that the $R_\mathrm{int}$ method (Sect. \ref{sec:rintyso}) is not afected by extragalactic contaminants. The H-band magnitude, which is required by the method, is more affected by extiction than mid-infrared bands. Specifically, our above established assumption of $\Delta A_{[4.5]}=1$ for extragalactic sources would be converted into $\Delta A_H \approx 3.6$ \citep[through][]{indebetouw+05}. This heavy H-band extinction is sufficient to make unobservable the relevant extragalactic contaminants (i.e. those that are faint enough not to be detectable in [8.0] or [5.8] despite their YSO-like colors).

\subsection{The benefits of $R_\mathrm{int}$ on the assessable sample} \label{sec:assessable}

In this section, we provide a detailed comparison between the point-source samples that can be evaluated by the $R_\mathrm{int}$ method (Sect. \ref{sec:rintyso}) and phase 2 of the G09 method. First, we note that all sources assessable by the G09 method can also be evaluated by our whole methodology (which is actually made of the adapted phase 1 plus the $R_\mathrm{int}$ method), but not vice versa.

Overall, our whole method is capable to evaluate 5048 sources in the OMEGA2000 field, a $\sim 36 \% $ gain relative to the 3718 from Sect. \ref{sec:photbias}. This increase majorly attributable to 915 $R_\mathrm{int}$ objects in the region lacking $4.5~\mu \mathrm{m}$ observations; therefore the gain is reduced to only 11\% when considering the remaining solid angle (i.e. in the NIR+MIR region). The superiority of the $R_\mathrm{int}$ method is stronger in the Cl05 region, where 75 objects are evaluated, a $\sim 29 \%$ gain (c.f. 58 objects from Sect. \ref{sec:photbias}). Hence it is clear that the $R_\mathrm{int}$ method helps mitigate the effects of the selection bias addressed in Sect. \ref{sec:photbias} to a significant extent.

If we only focus on the new 290 YSO candidates with  $R_\mathrm{int}>1.5$, the usefulness of the method is even clearer. Among these, at least 94 would have never been evaluated by the G09 method, owing to missing or inaccurate detection in some photometric band. Within the Cl05 region, the G09 method would be unable to assess 8 out of 34 new YSO candidates.

\section{Astrometric and photometric data} \label{sec:fulldata}

\begin{table*}
\caption{Astrometric data and infrared photometry for sources listed in Table \ref{tab:results}} \label{tab:fulldata}
\centering
\begin{tabular}{rcccccccc} 
\hline\hline
 Src & \multicolumn{2}{c}{J2000 coordinates} & $\varpi$ & $\varpi^\mathrm{corr}$ & $\mu_\alpha \cos \delta$ & {$\mu_\delta$} &  X-IR sep.\tablefootmark{a} & Epoch sep.\tablefootmark{b} \\
 \#  & 	R.A. 	 &	 Dec  	   & (mas) & (mas) & ($\mathrm{mas~yr^{-1}}$) & ($\mathrm{mas~yr^{-1}}$) &  (arcsec)  & (arcsec)   \\   
\hline
1  & 20:21:15.34 & +37:24:30.9 & $0.570 \pm 0.017$ & 0.598    & $-3.57 \pm 0.02$  & $-6.41 \pm 0.02$  & 0.35 &  0.15 \\ 
2  & 20:21:23.80 & +37:20:01.4 & $0.868 \pm 0.017$ & 0.910    & $0.64 \pm 0.02$   & $-4.84 \pm 0.02$  &      &       \\ 
3  & 20:21:56.20 & +37:21:28.5 & $1.649 \pm 0.012$ & 1.674    & $-6.11 \pm 0.01$  & $-20.33 \pm 0.01$ &      &       \\
4  & 20:21:23.80 & +37:20:01.4 & $0.253 \pm 0.030$ & 0.313    & $-3.83 \pm 0.03$  & $-6.05 \pm 0.03$  &      &       \\
5  & 20:21:35.54 & +37:23:29.6 & $1.772 \pm 0.012$ & 1.796    & $4.01 \pm 0.01$   & $4.06 \pm 0.01$   &      &       \\ 
6  & 20:21:34.04 & +37:23:53.1 & $0.537 \pm 0.026$ & 0.556    & $-3.87 \pm 0.03$  & $-7.66 \pm 0.03$  &      &       \\ 
7  & 20:21:46.38 & +37:15:37.0 & $0.669 \pm 0.017$ & 0.694    & $-1.55 \pm 0.02$  & $-4.52 \pm 0.02$  &      &       \\ 
8  & 20:21:50.52 & +37:28:57.0 & $0.909 \pm 0.018$ & 0.920    & $8.19 \pm 0.02$   & $2.29 \pm 0.02$   &      &       \\ 
9  & 20:21:38.66 & +37:25:15.0 & $0.563 \pm 0.011$ & 0.590    & $-3.52 \pm 0.01$  & $-5.96 \pm 0.01$  & 0.42 &  0.03 \\ 
10 & 20:21:33.56 & +37:24:51.4 & $0.581 \pm 0.020$ & 0.609    & $-3.42 \pm 0.02$  & $-5.95 \pm 0.02$  &      &       \\ 
11 & 20:21:37.02 & +37:24:17.1 & $-5.35 \pm 1.779$ & $-5.293$ & $-2.43 \pm 1.35$  & $-6.00 \pm 2.14$  &      &       \\ 
12 & 20:22:07.64 & +37:28:12.3 & $0.237 \pm 0.075$ & 0.274    & $-5.28 \pm 0.08$  & $-10.08 \pm 0.09$ &      &       \\ 
  ...  & ... & ...  &  ...  &  ...  &  ...  &  ...   &  ...  \\
\hline
\end{tabular}
\begin{tabular}{ccccccccc}
 \hline
  $J$	  &	$H$		&  $K$  	      &	[3.6]	&	[4.5]	&	[5.8]	&	[8.0]	& $f_{70 \mu \mathrm{m}}$ & $f_{160 \mu \mathrm{m}}$ \\
  (mag)	  &	(mag)		&  (mag)  	      &	(mag)	&	(mag)	&	(mag)	&	(mag)	&	(Jy)	 	&	(Jy)		     \\
 \hline
 $5.491 \pm 0.032$  &  $5.133 \pm 0.018$  &  $4.865 \pm 0.018$  &                   & $4.68 \pm 0.01$ & $4.30 \pm 0.01$ & $4.16 \pm 0.01$  &  &  \\
 $7.565 \pm 0.020$  &  $6.649 \pm 0.033$  &  $6.290 \pm 0.018$  &  $6.14 \pm 0.01$  &                 & $6.10 \pm 0.01$ &                  &  &  \\
 $7.643 \pm 0.020$  &  $7.019 \pm 0.018$  &  $6.886 \pm 0.021$  &  $6.81 \pm 0.01$  & $6.97 \pm 0.01$ & $6.83 \pm 0.01$ & $6.77 \pm 0.01$  &  &  \\
 $8.833 \pm 0.026$  &  $7.502 \pm 0.018$  &  $6.940 \pm 0.017$  &  $6.56 \pm 0.01$  & $6.67 \pm 0.01$ & $6.48 \pm 0.01$ & $6.45 \pm 0.01$  &  &  \\
 $7.708 \pm 0.020$  &  $7.224 \pm 0.016$  &  $7.107 \pm 0.015$  &  $7.05 \pm 0.01$  & $7.25 \pm 0.01$ & $7.06 \pm 0.01$ & $6.99 \pm 0.02$  &  &  \\
 $8.891 \pm 0.023$  &  $7.720 \pm 0.029$  &  $7.264 \pm 0.017$  &  $7.01 \pm 0.01$  & $7.12 \pm 0.01$ & $6.96 \pm 0.01$ & $6.91 \pm 0.02$  &  &  \\
 $8.321 \pm 0.019$  &  $7.626 \pm 0.017$  &  $7.334 \pm 0.020$  &  $7.09 \pm 0.01$  &                 & $7.08 \pm 0.01$ &                  &  &  \\
 $8.702 \pm 0.018$  &  $7.793 \pm 0.023$  &  $7.411 \pm 0.016$  &  $7.18 \pm 0.01$  & $7.28 \pm 0.01$ & $7.15 \pm 0.01$ & $7.13 \pm 0.02$  &  &  \\
 $7.878 \pm 0.017$  &  $7.616 \pm 0.023$  &  $7.455 \pm 0.016$  &  $7.35 \pm 0.02$  & $7.33 \pm 0.02$ & $7.43 \pm 0.08$ & $7.69 \pm 0.35$  &  &  \\
 $8.538 \pm 0.018$  &  $8.057 \pm 0.017$  &  $7.629 \pm 0.017$  &  $7.02 \pm 0.01$  & $6.78 \pm 0.01$ & $6.50 \pm 0.02$ & $6.15 \pm 0.04$  &  &  \\
 $12.198 \pm 0.026$ &  $9.244 \pm 0.017$  &  $7.811 \pm 0.014$  &  $6.69 \pm 0.01$  & $6.70 \pm 0.01$ & $6.33 \pm 0.01$ & $6.25 \pm 0.02$  &  &  \\
 $10.076 \pm 0.021$ &  $8.533 \pm 0.024$  &  $7.845 \pm 0.017$  &  $7.43 \pm 0.01$  & $7.52 \pm 0.01$ & $7.30 \pm 0.01$ & $7.27 \pm 0.01$  &  &  \\
...  &  ...  &   ...   &    ...   & ...  &  ...   & ... &   ...     &    ...    \\
\hline
\end{tabular}
\tablefoot{The full version of this table is available online at the CDS.\\
\tablefoottext{a}{Angular separation between the X-ray source and its infrared counterpart.}
\tablefoottext{b}{Angular separation between the 2009 and 2016 \textit{Chandra} detections.}
}
\end{table*}

Table \ref{tab:fulldata} shows the astrometric and infrared photometric data for all sources listed in Table \ref{tab:results}. Regarding X-rays, we only present astrometric information that is relevant to counterpart matching, while X-ray fluxes are postponed to the next paper of this series.

\end{appendix}

\end{document}